\documentclass[journal]{IEEEtran}
\makeatletter
\long\def\@makecaption#1#2{\ifx\@captype\@IEEEtablestring%
\footnotesize\begin{center}{\normalfont\footnotesize #1}\\
{\normalfont\footnotesize\scshape #2}\end{center}%
\@IEEEtablecaptionsepspace
\else
\@IEEEfigurecaptionsepspace
\setbox\@tempboxa\hbox{\normalfont\footnotesize {#1.}~~ #2}%
\ifdim \wd\@tempboxa >\hsize%
\setbox\@tempboxa\hbox{\normalfont\footnotesize {#1.}~~ }%
\parbox[t]{\hsize}{\normalfont\footnotesize \noindent\unhbox\@tempboxa#2}%
\else
\hbox to\hsize{\normalfont\footnotesize\hfil\box\@tempboxa\hfil}\fi\fi}
\usepackage[compatibility=false]{caption}
\DeclareCaptionFont{quackfont}{\fontfamily{ptm}\fontsize{7.5pt}{9pt}\selectfont}
\usepackage[labelformat=simple,font=quackfont]{subcaption}

\makeatother
\usepackage{lipsum}
\makeatletter
\let\MYcaption\@makecaption
\usepackage{cite}
\usepackage{graphicx}
\usepackage{subcaption}
\usepackage{float}
\usepackage{refstyle}
\usepackage{multicol, blindtext}
\usepackage{amsmath,amssymb,amsfonts}
\usepackage{adjustbox}
\newcommand{\norm}[1]{\left\lVert#1\right\rVert}

\usepackage[font=scriptsize]{caption}
\usepackage[font=footnotesize]{caption}
\usepackage{stfloats}
\usepackage{lmodern,bm}                
\usepackage[T1]{sansmath}
\usepackage{xcolor}
\usepackage{algorithm, algorithmic}
\usepackage{enumitem}
\usepackage{dutchcal}

\begin{document}

\title{An Unsupervised Deep Unfolding Framework for Robust Symbol Level Precoding}

\author{Abdullahi~Mohammad,~\IEEEmembership{Student Member,~IEEE,}
        Christos~Masouros,~\IEEEmembership{Senior Member,~IEEE,}
        and~Yiannis~Andreopoulos,~\IEEEmembership{Senior Member,~IEEE}}

\maketitle

\begin{abstract}
Symbol Level Precoding (SLP) has attracted significant research interest due to its ability to exploit interference for energy-efficient transmission. This paper proposes an unsupervised deep-neural network (DNN) based SLP framework. Instead of naively training a DNN architecture for SLP without considering the specifics of the optimization objective of the SLP domain, our proposal unfolds a power minimization SLP formulation based on the interior point method (IPM) proximal \textit{`log'} barrier function. Furthermore, we extend our proposal to a robust precoding design under channel state information (CSI) uncertainty. The results show that our proposed learning framework provides near-optimal performance while reducing the computational cost from $\mathbcal{O}(n^{7.5})$ to $\mathbcal{O}(n^{3})$ for the symmetrical system case where $n=\text{number of transmit antennas}=\text{number of users}$. This significant complexity reduction is also reflected in a proportional decrease in the proposed approach's execution time compared to the SLP optimization-based solution.
\end{abstract}


\IEEEpeerreviewmaketitle

\section{Introduction}
\IEEEPARstart{I}{nterference} has been known to yield a decrease in the throughput and communication reliability of a downlink multi-user multiple-input single-output (MU-MISO) wireless system. Traditionally, interference is regarded as the limiting factor against the ever-increasing needs for transmission rates and quality of service (QoS) in fifth-generation (5G) wireless communication systems and beyond \cite{masouros2011interference,masouros2015exploiting,masouros2018harvesting}. However, recent studies on interference exploitation have transformed the traditional paradigm in which known inferences are effectively managed \cite{masouros2011interference,masouros2015exploiting,ni2015beamforming,masouros2018harvesting,hong2014applications}. Consequently, transmit beamforming techniques for the downlink channels for power minimization problems under specific QoS become imperative for high-throughput systems under interference.\par 
The idea of exploiting interference was first introduced by Masouros and Alsusa \cite{masouros2007novel}, where instantaneous interference was classified into constructive and destructive. Initial suboptimal approaches to exploit constructive interference (CI) were first introduced by Masouros \textit{et al.} \cite{4801492}\cite{masouros2010correlation}. The first form of optimization-based CI precoding was introduced in the context of vector perturbation precoding through a quadratic optimization approach \cite{masouros2014vector}. A convex optimization-based CI scheme termed symbol-level-precoding technique was proposed first with strict phase constraints on the received constellation point \cite{alodeh2015constructive}, and with a robust relaxed-angle formulation \cite{masouros2015exploiting}. We refer to recent work \cite{masouros2014vector, alodeh2015constructive,amadori2016constant,alodeh2017symbol,li2018massive} for more details on the optimization-based CI precoding techniques.\par
As a result of the performance gains over conventional block-level-precoding (BLP) schemes, the idea of CI has been applied in many domains, such as vector perturbation \cite{li2015two}, wireless information and power transfer \cite{timotheou2016exploiting}, mutual coupling exploitation \cite{li2017exploiting}, multiuser MISO downlink channel \cite{spano2017symbol}, directional modulation \cite{kalantari2017spatial}, relay and cognitive radio \cite{masouros2011interference,law2017transmit}. Despite the superior performance offered by CI-based precoding methods, their increased computational complexity can hinder their practical application when performed on a symbol-by-symbol basis. To address this, Li and Masouros \cite{li2018interference} proposed an iterative closed-form precoding design with optimal performance for CI exploitation in the MISO downlink by driving the optimal precoder's mathematical Lagrangian expression and Karush–Kuhn–Tucker conditions for optimization with both strict and relaxed phase rotations.\par
Lately, there is growing interest in using deep neural networks (DNNs) for wireless physical layer design \cite{mohammad2020complexity,alkhateeb2018deep,mohammad2020accelerated}. More relevant to this work are the learning-based precoding schemes for MU-MISO downlink transmission \cite{huang2018unsupervised,sun2018learning,xia2019deep,huang2019fast,lei2021ci}. The benefit of using DNNs is that the computational burden of the learning algorithm can be controlled via online training, and a variety of loss functions can be used for each optimization objective. One of the earliest attempts of using DNNs models for beamforming design was the work of Alkhateeb \textit{et al.} \cite{alkhateeb2018deep}, where a learning-based coordinated beamforming technique was proposed for link reliability and frequent poor hand-off between base stations (BSs) in millimeter-wave (mmWave) communications. Kerret and Gesbert \cite{de2018robust} introduced DNNs precoding scheme to address the \textit{``Team Decision problems"} for a decentralized decision making in multiple-input-multiple-output (MIMO) settings. Huang \textit{et al.} \cite{huang2018unsupervised} proposed a fast beamforming design based on unsupervised learning that yielded performance close to that of the weighted minimum mean-squared error (WMMSE) algorithm. A DNN-based precoding strategy that utilized a heuristic solution structure of the downlink beamforming was proposed by Huang \textit{et al.} \cite{huang2019fast}. Furthermore, Xia \textit{et al.} \cite{xia2019deep} developed deep convolutional neural networks (CNNs) framework for downlink beamforming optimization. The framework exploits expert knowledge based on the known structure of optimal iterative solutions for sum-rate maximization, power minimization, and SINR balancing problems.\par 
DNN methods are typically used for unconstrained optimization problems. Therefore, most of the DNN-based strategies for wireless physical layer designs are based on supervised learning to approximate the optimal solutions. Using such approaches, the constraints are implicitly contained in the training dataset obtained from conventional optimization solutions. However, if obtaining optimal solutions via traditional optimization methods is very computationally expensive (or infeasible), using supervised learning methods for DNN-based method may not be practical.\par
Furthermore, the common approach for solving constrained optimization with DNN for wireless physical layer design is via function approximation. It involves solving the problem, first using iterative algorithms or convex optimization techniques, and finally approximating the optimal solution with a DNN architecture \cite{sun2018learning,huang2019fast,xia2019deep}. Accordingly, the major drawback of these proposals is that the efficacy of supervised learning is bounded by the assumptions and accuracy of the optimal solutions obtained from the structural optimization algorithm.\par
This work proposes an unsupervised learning-based approach for precoding design by exploiting known interference in MU-MISO systems for the power minimization problem under SINR constraints. The learning framework is designed by unfolding an interior point method (IPM) iterative algorithm via \textit{`log'} barrier function. The proposed learning-based precoding scheme does not require generating the training dataset from the conventional optimization solutions, thereby saving considerable computational effort and time. Our contributions are summarized below:
\begin{itemize}
    \item We introduce an unsupervised DNN-based power minimization SLP scheme for MU-MISO downlink transmission. The proposed framework is designed by unfolding an IPM algorithm via a \textit{`log'} barrier function that exploits the convexity associated with the SLP inequality constraints. The learning framework utilizes the domain knowledge to derive the Lagrange function of the original SLP optimization as a loss function. This is used to train the network in an unsupervised mode to learn a set of Lagrangian multipliers that directly minimize the objective function to satisfy the constraints. A regularization parameter is added to the Lagrange function to aid the training convergence, and we provide detailed formulations leading to the unfolded unsupervised learning architecture for constrained optimization problems. 
    \item We extend the formulation to design a robust learning-based precoder where the uncertainty in channel estimation is considered.
    \item We derive analytic expressions for the computational complexity of various SLP and the proposed unsupervised learning precoding schemes. Our analysis demonstrates that the proposed deep unfolding (DU) framework offers a theoretical, computational complexity reduction from $\mathbcal{O}(n^{7.5})$ to $\mathbcal{O}(n^{3})$ for the symmetrical system case where $n$ = number of transmit antennas = number of users This is reflected in a commensurate decrease in the execution time as compared to the SLP optimization-based method.   
\end{itemize}\par
The remainder of the paper is organized as follows: The system model and the methods for traditional precoding and SLP optimization-based for downlink MU-MISO system are presented in Section \ref{section2}. The proposed unsupervised DU-based precoding designs under perfect channel condition for power minimization are introduced in Section \ref{non_robust_prob} and extension to a robust precoding design under uncertainty channel condition is described in Section \ref{robust_problem}. Section \ref{data_train_complexity} presents detailed analytic computational complexity evaluation of the proposed precoding schemes. Simulations and results are presented in Section \ref{results}. Finally, Section \ref{conclusion} summarizes and concludes the paper.\par
{\textbf{Notations:} We use bold uppercase symbols for matrices, bold lowercase symbols for vectors and lowercase symbols for scalars. The $\mathbcal{l}_{2}$-\text{norm} and $\mathbcal{l}_{1}$-\text{norm} are denoted by $\norm{\cdot}_{2}$ and $\norm{\cdot}_{1}$, respectively. The $|\cdot|$ represents the absolute value and $\boldsymbol{\theta}_{i}$ is the \textit{i}-th trainable parameter associated with DNN layers. Operators Re$(\cdot)$ and Im$(\cdot)$ represent real and imaginary parts of a complex vector, respectively. Finally, notations $\mathbcal{L}(\cdot)$ and $\mathbcal{H}(\cdot)$ are reserved for the loss and parameter update functions, respectively.}    
\section{System Model and Problem Description}\label{section2}
\subsection{Conventional Block Level Precoding for Power Minimization} 
Consider a signle-cell downlink channel with $N_{t}$ transmit antennas at the BS transmitting to $K$ single-antenna users. Assume a quasi-static flat-fading channel between the BS and the users, denoted by $\mathbf{h}_{i}\in \mathbb{C}^{{N}_{t}\times1}$. The received signal at user $i$ is given by
\begin{equation}\label{received_signal}
\begin{split}
y_{i}&=\mathbf{h}^{T}_{i}\sum\limits ^{K}_{k=1}\mathbf{w}_{k}{s}_{k}+v_{i}\\
&=\mathbf{h}^{T}_{i}\sum_{k=1}^{K}\mathbf{w}_{k}e^{j(\varphi_{k}-\varphi_{i})}{s}_{i}+v_{i}
\end{split}
\end{equation}
where $\mathbf{h}_{i}$, $\mathbf{w}_{i}$, ${s}_{i}$, ${v}_{i}$ and $\varphi_{i}$ represent the channel vector, precoding vector, data symbol, received noise and phase rotation for the \textit{i}-th user. Conventionally, the power minimization problem seeks to minimize the average transmit power by treating all interference as detrimental subject to QoS constraints as defined below \cite{bjornson2014optimal}
\begin{equation} \label{conv_power_min}
    \begin{aligned}
    & \underset{\mathbf{\{w_{i}\}}}{\text{min}}
    & & {\sum_{i=1}^K\norm{ \mathbf{w}_{k}}_{2}^2} \\
    & \text{s.t.}
    & &  \frac{|\mathbf{h}_{i}^{T}\mathbf{w}_{i}|^{2}}{\sum_{k=1,k\neq i}|\mathbf{h}_{i}^{T}\mathbf{w}_{k}|^{2}+{v}_{0}}\geq\Gamma_{i}\ , \ \forall {i}.
    \end{aligned}
\end{equation}
where $\Gamma_{i}$ is the SINR threshold of the \textit{i}-th user.
It has been proven that problem (\ref{conv_power_min}) is suboptimal from an instantaneous point of view, as it does not take into account the fact that interference can constructively enhance the received signal power \cite{masouros2010correlation}.
\begin{figure}
    \includegraphics[width=3.2in,height=2.5in]{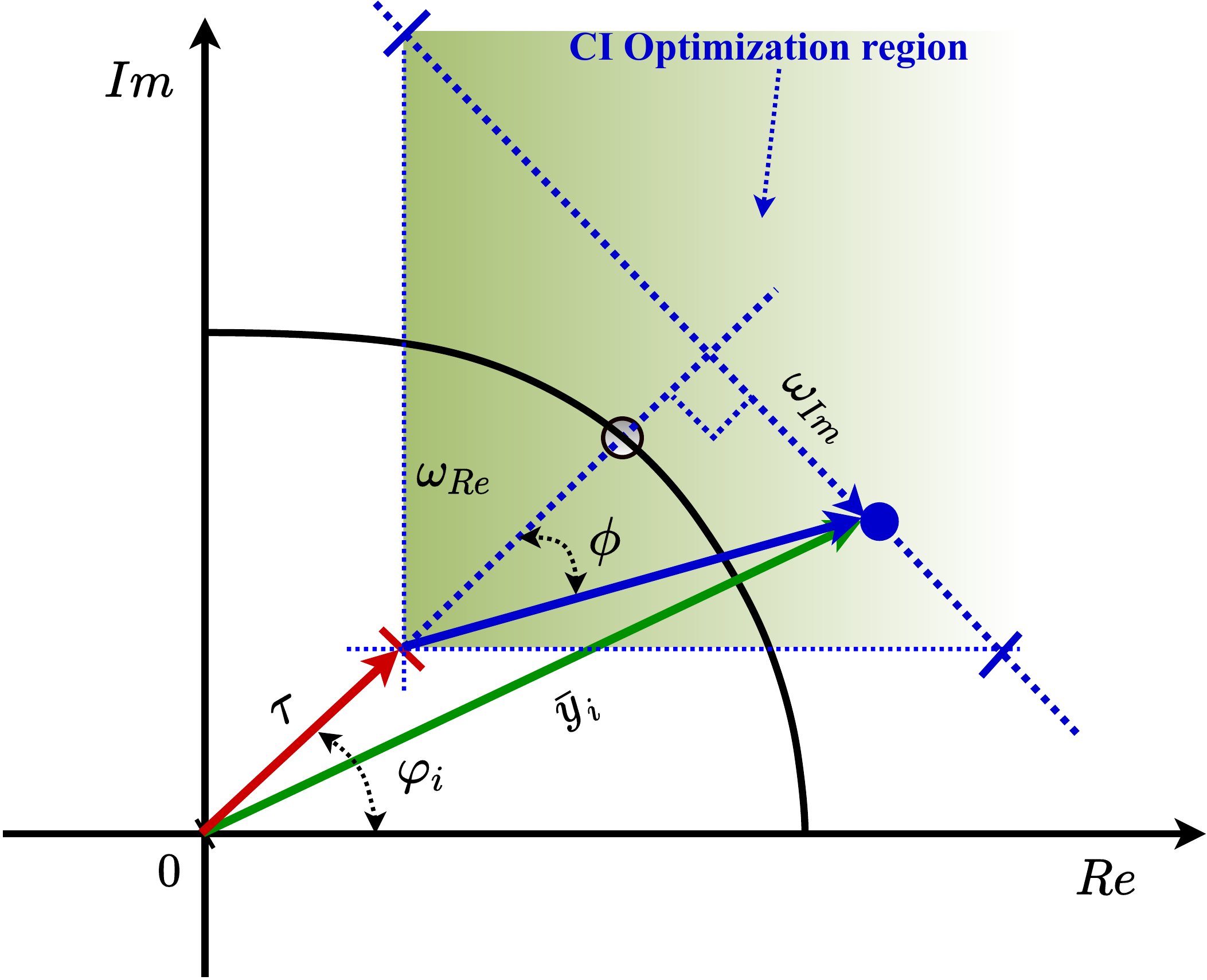}
\caption{Generic geometrical optimization regions for interference exploitation for Precoding design in $\mathbb{M}$-PSK \cite{masouros2015exploiting}.}
\label{fig:CI_GRAHICAL_REP}
\end{figure}
\subsection{Power Minimization via Symbol-Level Precoding}
With the aim of utilizing the instantaneous interference in a multi-user downlink channel scenario, the interference can be categorized into constructive and destructive based on the known standards described \cite{masouros2013known}.
Based on this, CI is defined as the interference that nudges the received symbols off the modulated-symbol constellation's decision thresholds \cite{masouros2015exploiting}. Fig. \ref{fig:CI_GRAHICAL_REP} shows the generic geometrical representation of the CI, where the received signal is expressed as $\bar{\mathbf{y}}_{i} \triangleq \mathbf{h}_{i}^{T}\sum_{k=1}^{K}\mathbf{w}_{k}{e}^{j(\varphi_{k}-\varphi_{1})}$. From the received symbol expression, the real and imaginary parts are respectively given by: $\boldsymbol{\omega}_{Re}=\text{Re}({\bar{\mathbf{y}}_{i}})$ and $\boldsymbol{\omega}_{Im}=\text{Im}({\bar{\mathbf{y}}_{i}})$. In Fig. \ref{fig:CI_GRAHICAL_REP}, we show an indicative example corresponding to the constellation point $1+j$ in the QPSK constellation, where the green shaded area represents the constructive region of the constellation, based on the minimum distance $(\tau)$ from the decision boundaries. The value of $\tau$ is determined by the SNR constraints. In line with the preceding description, problem (\ref{conv_power_min}) is adjusted to include CI in the power minimization formulation. This allows the interfering signals to align with the symbol of interest constructively through precoding, contributing to the desired signal. Therefore, for $\mathbb{M}$-PSK, the power minimization SLP can be reformulated based on the constructive/destructive interference classification criteria \cite{li2020tutorial}. The maximum phase shift in the CI region is given by $\phi=\pm\frac{\pi}{\mathbb{M}}$, where $\mathbb{M}$ is the modulation order. Therefore, the SLP optimization is given by \cite{masouros2015exploiting} 
\begin{equation} \label{CI_optimization}
    \resizebox{1.01\hsize}{!}{$%
    \begin{aligned}
    & \underset{\mathbf{\{w_{i}\}}}{\text{min}}
    & & {\norm{\sum_{k=1}^K \mathbf{w}_{k}{e}^{j(\varphi_{k}-\varphi_{1})} }_{2}^2} \\
    & \text{s.t.}
    & & \left|\text{Im}{\left(\mathbf{h}_{i}^{T}\sum_{k=1}^{K}\mathbf{w}_{k}e^{j(\varphi_{k}-\varphi_{i})}\right)}\right|\leq\\&&& \left(\text{Re}{\left(\mathbf{h}_{i}^{T}\sum_{k=1}^{K}\mathbf{w}_{k}e^{j(\varphi_{k}-\varphi_{i})}\right)- \sqrt{\Gamma_{i}v_{0}}}\right)\text{tan}\phi\ ,\ \forall {i}.
    \end{aligned}$}
\end{equation}
\section{Learning-Based Power minimization for SLP}\label{non_robust_prob}
This section presents the formulation of a learning-based CI power minimization problem for SLP. Throughout this section, we assume a perfect CSI known at the BS.\par
Motivated by the recent adoption of an IPM for image restoration \cite{bertocchi2020deep}, we propose an unsupervised learning framework that unfolds a constrained optimization problem into a sequence of learning layers/iterations for a multi-user MISO beamforming. We first convert (\ref{CI_optimization}) into a standard IPM formulation containing a slack variable, where necessary. The measure of the fidelity of the solution to (\ref{CI_optimization}) is determined by learning a set of penalty parameters in the form of Lagrange multipliers associated with the constraints. From (\ref{CI_optimization}), we define the following \begin{equation} \label{h_hat}
     \Hat{\mathbf{h}}_{i}=\mathbf{h}_{i}\sum_{k=1}^{K}e^{j(\varphi_{k}-\varphi_{i})}
\end{equation}  
 \begin{equation} \label{w}
     {\mathbf{w}}=\sum_{k=1}^{K}\mathbf{w}_{k}.
\end{equation}\par
Accordingly, to ease the analysis, we partition the complex rotations into the real and imaginary parts as follows \begin{subequations}\label{H_w}
\begin{align}
\label{H_hati}
\Hat{\mathbf{h}}_{i} & = \Hat{\mathbf{h}}_{Ri}+j\Hat{\mathbf{h}}_{Ii}\\
\label{w_ir}
\mathbf{w} & =\mathbf{w}_{R}+j\mathbf{w}_{I}
\end{align}
\end{subequations}
where $\Hat{\mathbf{h}}_{Ri}=\text{Re}(\Hat{\mathbf{h}}_{i})$, $\Hat{\mathbf{h}}_{Ii}=\text{Im}(\Hat{\mathbf{h}}_{i})$, $\mathbf{w}_{R}=\text{Re}(\mathbf{w})$ and $\mathbf{w}_{I}=\text{Im}(\mathbf{w})$.\par 
The product of complex rotations of (\ref{H_hati}) and (\ref{w_ir}) can be written as 
\begin{equation}\label{pro_comp}
    \Hat{\mathbf{h}}_{i}\mathbf{w}=(\Hat{\mathbf{h}}_{Ri}+j\Hat{\mathbf{h}}_{Ii})(\mathbf{w}_{R}+j\mathbf{w}_{I}).
\end{equation}

Using (\ref{H_w}), the real and imaginary parts of (\ref{pro_comp}) can be written in vector forms as follows
\begin{subequations}\label{Hw_vec}
\begin{align}
\label{Hw_Re}
\text{Re}(\Hat{\mathbf{h}}_{i}\mathbf{w})&=\begin{bmatrix}
\Hat{\mathbf{h}}_{Ri} \ \ \Hat{\mathbf{h}}_{Ii}\
\end{bmatrix}
\begin{bmatrix}
\mathbf{w}_{R}\\ -\mathbf{w}_{I}\
\end{bmatrix} \\
\label{Hw_Im}
\text{Im}(\Hat{\mathbf{h}}_{i}\mathbf{w})&=\begin{bmatrix}
\Hat{\mathbf{h}}_{Ri} \ \ \Hat{\mathbf{h}}_{Ii}\
\end{bmatrix}
\begin{bmatrix}
\mathbf{w}_{I}\\ \mathbf{w}_{R}\
\end{bmatrix}
\end{align}
\end{subequations} 

Let $\boldsymbol{\Lambda}=[\Hat{\mathbf{h}}_{Ri}\ \ \Hat{\mathbf{h}}_{Ii}]^{T}
$, $\mathbf{w}_{1} = [
\mathbf{w}_{R}\ -\mathbf{w}_{I}]^{T}$ and $\mathbf{w}_{2} = [
\mathbf{w}_{I}\ \ \mathbf{w}_{R}]^{T}$
\begin{equation}
\text{Re}(\Hat{\mathbf{h}}_{i}^{T}\mathbf{w})=\boldsymbol{\Lambda}_{i}^{T}\mathbf{w}_{1}\ \text{and}\  \text{Im}(\Hat{\mathbf{h}}_{i}^{T}\mathbf{w}) = \boldsymbol{\Lambda}_{i}^{T}\boldsymbol{\Pi}\mathbf{w}_{1}
\end{equation}
where \begin{equation}
{\mathbf{w}_{2}=\boldsymbol{\Pi}}\mathbf{w}_{1}\ \text{and}\ \boldsymbol{\Pi}=\begin{bmatrix}
\mathbf{O}_{N_{t}} & -\mathbf{I}_{N_{t}}\\
\mathbf{I}_{N_{t}} & \mathbf{O}_{N_{t}}
\end{bmatrix}; \ \in \mathbb{R}^{2N_{t} \times 2N_{t}},
\end{equation}
Note that $\mathbf{I}_{N_{t}}$ is the identity matrix and $\mathbf{O}_{N_{t}}$ the matrix of zeros, respectively. Using the above definitions, problem (\ref{CI_optimization}) can be recast into its mutlicast formulation \cite{masouros2015exploiting}
\begin{equation} \label{P_relaxed1}
    \begin{aligned}
    & \underset{\mathbf{\{w_{1}\}}}{\text{min}}
    & & {\norm{\mathbf{w}_{1}}_{2}^2} \\
    & \text{s.t.}
    & & \left|{\boldsymbol{\Lambda}_{i}^{T}\boldsymbol{\Pi}\mathbf{w}_{1}}\right|\leq\left(\boldsymbol{\Lambda}_{i}^{T}\mathbf{w}_{1}-\sqrt{\Gamma_{i}v_{0}}\right)\text{tan}\phi\ ,\ \forall {i}
    \end{aligned}
    \end{equation}
\subsection{Interior Point Method}\label{IMP}
Consider a general form of a nonlinear constrained optimization of the form \cite{hauser2007interior}
\begin{equation}
    \begin{aligned}
    & \underset{\mathbf{x \in{\mathbb{R}^{N}}}}{\text{min}}
    & & {f(\mathbf{x})} \\
    & \text{s.t.}
   && g(\mathbf{x})\geq0\\
   &&&\left.\begin{aligned} 
    C(\mathbf{x})=0
    \end{aligned}
    \right.,
    \end{aligned}
\end{equation}
The rationale of adopting IPM is to substitute the initial constrained optimization problem by a chain of unconstrained sub-problems of the form
\begin{equation}\label{sub_prob0}
    \begin{aligned}
    & \underset{\mathbf{x \in{\mathbb{R}^{N}}}}{\text{min}} f(\mathbf{x})+\lambda{{C(\mathbf{x})}}+\mu{{B(\mathbf{x})}}.
    \end{aligned}
\end{equation}
where ${B}(\cdot)\triangleq-\sum\ln{(\cdot)}$ is the logarithmic barrier function associated with inequality constraint with unbounded derivative at the boundary of the feasible domain, ${C}(\cdot)$ is a function associated with equality constraint, $\mu$ and $\lambda$ are the Lagrangian multipliers for inequality and equality constraints, respectively. For $K$ users, we define a vector $\boldsymbol{\mu}\triangleq[\mu_{1},\cdots,\mu_{K}]$.\par
Following the above line of argument, the unconstrained sequence of (\ref{P_relaxed1}) per user can be written as
\begin{equation}\label{sub_prob}
    \begin{aligned}
    & \underset{\mathbf{w \in{\mathbb{R}^{2N_{t}\times1}}}}{\text{min}} f(\mathbf{w}_{1})+{\mu}{{B(\mathbf{w}_{1})}},
    \end{aligned}
\end{equation}
To facilitate the solution of (\ref{P_relaxed1}) , we introduce additional notations. For every inequality constraint, $\gamma \in \{0,+\infty\}$ and $\mathbf{w}_{1} \in \mathbb{R}^{2N_{t}\times1}$, we define the proximity function as in \cite{hauser2007interior} with respect to (\ref{sub_prob}), which we shall later use to compute the projected gradient descent as 
\begin{equation}\label{prox_op}
   \begin{aligned}
   \text{prox}_{\gamma{\mu}{B}}{(\mathbf{w}_{1})}= & \underset{\mathbf{w_{1} \in{\mathbb{R}^{2N_{t}\times{1}}}}}{\text{argmin}}
    & & {\frac{1}{2}\norm{\mathbf{w}_{0}-\mathbf{w}_{1}}}_{2}^{2}+\gamma{\mu}{B}({\mathbf{w}_{1}}), 
    \end{aligned}
\end{equation}
where $\gamma$ is the step-size for computing the gradients and $\mathbf{w}_{0}$ is the initial value of the precoding vector. To convert (3) into its equivalent barrier function problem, we integrate the inequality constraint into the objective by translating it into a barrier term as follows \cite{pustelnik2017proximity}
\begin{equation}\label{b2}
    \begin{aligned}
    & \underset{\mathbf{w_{1}}}{\text{min}}
    & & {f(\mathbf{w}_{1})}-{\mu}\sum_{i=1}^{p}{\ln{\left(g({\mathbf{w}}_{1i})\right)}} \\
    & \text{s.t.}
    & & {{C(\mathbf{w}_{1})}=0}
    \end{aligned}
\end{equation}
where $g({\mathbf{w}}_{1})=\left(\boldsymbol{\Lambda}_{i}^{T}\mathbf{w}_{1}-\sqrt{\Gamma_{i}v_{0}}\right)\text{tan}\phi-\left|{\boldsymbol{\Lambda}_{i}^{T}\boldsymbol{\Pi}\mathbf{w}_{1}}\right|$ and $p$ is the number of the optimization variables.\par 
Going back to our initial SPL optimization to apply this framework, first we rewrite the constraint of (\ref{P_relaxed1}) as
\begin{equation}\label{hyperslab_const}
    {a}\leq {\boldsymbol{\Lambda}_{i}^{T}\boldsymbol{\Pi}\mathbf{w}_{1}}\leq{b},
\end{equation}
where
\begin{subequations}
\begin{align}
    {a} & =-\left(\boldsymbol{\Lambda}_{i}^{T}\boldsymbol{\Pi}\mathbf{w}_{1}-\sqrt{\Gamma_{i}v_{0}}\right){\text{tan}{\phi}},\\
    {b} & =\left(\boldsymbol{\Lambda}_{i}^{T}\boldsymbol{\Pi}\mathbf{w}_{1}-\sqrt{\Gamma_{i}v_{0}}\right){\text{tan}{\phi}}.
\end{align}
\end{subequations}

Therefore, the original problem (\ref{P_relaxed1}) becomes
\begin{equation}\label{P_relaxed2}
    \begin{aligned}
    & \underset{\mathbf{\{w_{1}\}}}{\text{min}}
    & & {\norm{\mathbf{w}_{1}}_{2}^2} \\
    & \text{s.t.}
    & & {a} \leq {\boldsymbol{\Lambda}_{i}^{T}\boldsymbol{\Pi}\mathbf{w}_{1}}\leq{b}\ ,\ \forall {i}.
    \end{aligned}
\end{equation}
It is apparent that the constraint of (\ref{P_relaxed2}) is contained within a hyperslab \cite{boyd2004convex}.
\subsubsection{Hyperslab Constraints}\label{hyperslab}
Given the constraint in (\ref{P_relaxed2}), the precoding vector $\mathbf{w}_{1}$ is contained within a set of hyperslab $\mathcal{C}$ and also bounded by $\{{a},\ {b}\}$. Therefore, $\mathcal{C}$ is defined as follows
\begin{equation}\label{hypeslab}
\mathbcal{C}=\{\mathbf{w}_{1}\in \mathbb{R}^{N_{t}\times1}\}|_{{a} \leq \boldsymbol{\Lambda}^{T}\boldsymbol{\Pi}\mathbf{w}_{1}\leq{b}}.
\end{equation}
For all $\gamma > 0$ and ${\mu}>0$, a proximity barrier function related to (\ref{hypeslab}) is given by
\begin{equation}\label{prox_func_rel2}
    \resizebox{1.02\hsize}{!}{$\begin{array}{ll}
    {B}({\mathbf{w}}_{1}) = \left \{
    \begin{aligned}
    &-\ln{\left({b}-\Hat{{w}}_{1}\right)}-\ln{\left(-{a}+\Hat{{w}}_{1}\right)},&& \text{if}\ -{a} \leq \Hat{{w}}_{1}\leq{b} \\
    &+\infty, && \text{otherwise}
    \end{aligned} \right.
    \end{array}$}
\end{equation} 
where for convenience, we let $\Hat{{w}}_{1}=\boldsymbol{\Lambda}_{i}^{T}{\boldsymbol{\Pi}}{\mathbf{w}_{1}}$.

\subsection{Proximity Operator for the SLP Formulation}\label{Prox_Operator_relaxed}
To unfold (\ref{P_relaxed2}) into learning framework using IPM, we use its equivalent proximity \textit{`log'} barrier function (\ref{prox_func_rel2}) and the proximal operator of $\gamma {\mu} B(\mathbf{w}_1)$ for every $\mathbf{w}_{1}$ defined as
\begin{equation}\label{prox_opr_rl2}
     \Phi(\mathbf{w}_{1},\gamma,{\mu})=\text{prox}_{\gamma\boldsymbol{\mu}{{B}}}{(\mathbf{w}_{1})}=\mathbf{w}_{1}+\frac{X(\mathbf{w}_{1},\gamma,{\mu})-\boldsymbol{\Lambda}_{i}^{T}\mathbf{w}_{1}}{\| \boldsymbol{\Lambda}_{i}\|_{2}^2}{\boldsymbol{\Lambda}_{i}}
\end{equation}
where ${X}$ is a typical solution of the following cubic equation of the form
\begin{multline}\label{cubic_equation}
   {x}^{3} -\left({b}+{a} +\boldsymbol{\Lambda}^{T}\mathbf{w}_{1}\right){x}^{2} +\\ \left({b}{a}+\boldsymbol{\Lambda}^{T}\mathbf{w}_{1}({b}+{a})-2\gamma{{\mu}} \| \boldsymbol{\Lambda}\|_{2} ^{2}\right){x}\\
   +\left(-{b}{a}\boldsymbol{\Lambda}^{T}\mathbf{w}_{1}+\gamma{\mu} \left({b}+{a} \right)\| \boldsymbol{\Lambda}\|_{2} ^{2}\right)=0.
\end{multline} 
It is important to note that the solution to (\ref{cubic_equation}) is obtained using the analytic solution of the cubic equation.
To build the structure of the learning framework, as detailed in \cite{bertocchi2020deep}, we need to obtain the Jacobian matrix of $\Phi(\mathbf{w}_{1},\gamma,{\mu})$ with respect to $\mathbf{w}_{1}$ and the derivatives with respect to $\gamma$ and $\mu$ as follows
\begin{multline}\label{jacob_mat_relax}
\mathcal{J}_{{\Phi}}\mid_{(\mathbf{w}_{1})}=\mathbf{I}_{2N_{t}}+\frac{1}{\norm{\boldsymbol{\Lambda}_{i}}_{2}^{2}} \ \times \\ \left(\frac{\left({b}-X(\mathbf{w}_{1},\gamma,{\mu})\right)\left({a}-X(\mathbf{w}_{1},\gamma,{\mu})\right)}{\Upsilon(\mathbf{w}_{1},\gamma,{\mu})}-1\right)\boldsymbol{\Lambda_{i}\Lambda_{i}^{T}}
\end{multline} 

\begin{equation}\label{deriv_mu_rl}
    \Delta_{{\Phi}}\mid_{({{\mu}})}=\frac{{-\gamma}\left({b}+{a}-2X(\mathbf{w}_{1},\gamma,{\mu})\right)}{\Upsilon(\mathbf{w}_{1},\gamma,{\mu})}\boldsymbol{\Lambda_{i}}
\end{equation}

\begin{equation}\label{deriv_gamma_rl}
    \Delta_{{\Phi}}\mid_{({\gamma})}=\frac{{-{\mu}}\left({b}+{a}-2X(\mathbf{w}_{1},\gamma,{\mu})\right)}{\Upsilon(\mathbf{w}_{1},\gamma,{\mu})}\boldsymbol{\Lambda_{i}},
\end{equation}
where $\mathbf{I}_{2N_{t}} \in\mathbb{R}^{2N_{t}\times2N_{t}}$. For hyperslab constraints, $\boldsymbol{\Upsilon}(\cdotp)$ is the derivative of (\ref{cubic_equation}) with respect to ${x}$. Finally, using similar abstraction as in subsection \ref{IMP}, the SLP formulation can be expressed as a succession of sub-problems with respect to the inequality constraint
\begin{equation}\label{prox_func_relaxed}
    \begin{aligned}
    & \underset{\mathbf{w_{1} \in{\mathbb{R}}^{2N_{t}\times1}}}{\text{min}}
    & & {\norm{\mathbf{w}_{1}}}_{2}^{2}+\lambda{\mathbf{w}_{1}} +{\mu}{{B}(\mathbf{w}_{1})}.
    \end{aligned}
\end{equation}

It is important to note that the original problem (\ref{P_relaxed1}) does not have an equality constraint. However, the term $\lambda\mathbf{w}_{1}$ introduced in (\ref{prox_func_relaxed}) is to provide additional stability to the network. Using the proximity operator of the barrier, the update rule for every iteration is given by
\begin{equation}\label{beam_update_relaxed}
\mathbf{w}_{1}^{[r+1]}=\text{prox}_{\gamma^{[r]}\boldsymbol{\mu}^{[r]}B}\left(\mathbf{w}_{1}^{[r]}-\gamma^{[r]}\Delta{{D}(\mathbf{w}_{1}^{[r]},\lambda^{[r]})}\right)
\end{equation}
where 
\begin{equation}\label{eq_constr_rl}
    {D}(\mathbf{w}_{1}^{[r]},\lambda^{[r]})={\norm{\mathbf{w}_{1}}}_{2}^{2}+\lambda{\mathbf{w}_{1}},
\end{equation}
and $\Delta=\frac{\partial{{D}(\mathbf{w}_{1}^{[r]},\lambda^{[r]})}}{\partial{\mathbf{w}_{1}^{[r]}}}$.

\subsection{Deep SLP Network (SLP-DNet)}
\begin{figure*}
    \centering
    \includegraphics[width=7.3in,height=3.3in]{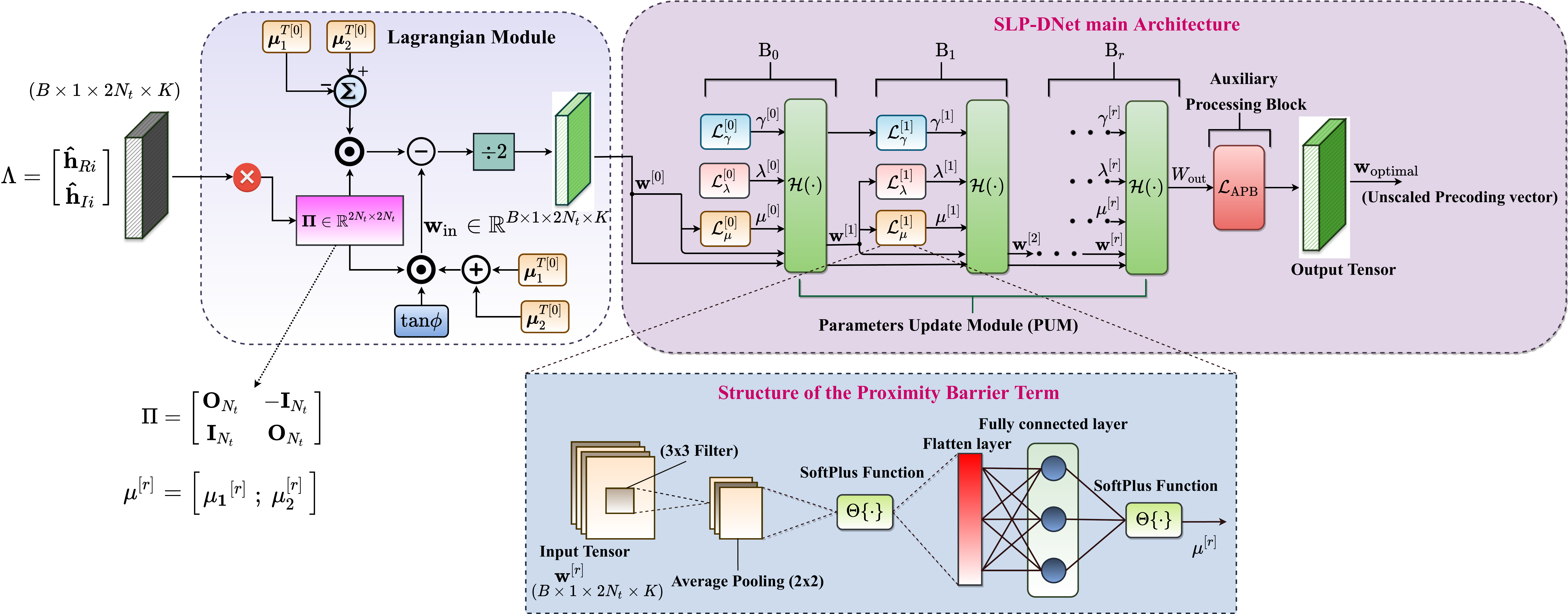}
    \caption{Complete SLP-DNet Architecture, showing the parameter update mouule, the auxiliary processing block.}
    \label{fig:DNBF_Arch}
\end{figure*}

To build the proposed learning-based SLP architecture, we combine an IPM with a proximal forward-backward procedure as shown in Algorithm \ref{algorithm_1} and transform it into an NN structure represented by the proximity barrier term (see Fig. \ref{fig:DNBF_Arch}). The learning architecture strictly follows the formulation (\ref{beam_update_relaxed}). We show a striking similarity between our proposal and the feed-forward. Intuitively, we form cascade layers of NN from (\ref{beam_update_relaxed}) as follows
\begin{equation}\label{beam_update2}
\mathbf{w}_{1}^{[r+1]}=\text{prox}_{\gamma^{[r]}{\mu}^{[r]}{B}}\left[\left(\mathbf{I}_{N_{t}}-2\gamma^{[r]}\right)\mathbf{w}_{1}^{[r]}+\gamma^{[r]}\lambda^{[r]}\textbf{1}^{T}\right].
\end{equation}
where $\textbf{1}\in\mathbb{R}^{1\times2N_{t}}$ is a vector of ones. By letting $\mathbf{W}_{r}=\mathbf{I}_{2N_{t}}-2\gamma^{[r]}$, $\mathbf{b}_{r}=\gamma^{[r]}\lambda^{[r]}\textbf{1}^{T}$ and $\boldsymbol{\Theta}_{r}=\text{prox}_{\gamma^{[r]}{\mu}^{[r]}{B}}$, the \textit{r}-layer network $\mathcal{L}^{[r-1]}\cdots \mathcal{L}^{[0]}$ will correspond to the following
\begin{multline}\label{neural_net}
  \boldsymbol{\Theta}_{0}\left(\mathbf{W}_{0}+\mathbf{b}_{0}\right),\cdots,\boldsymbol{\Theta}_{r}\left(\mathbf{W}_{r}+\mathbf{b}_{r}\right)
\end{multline}
where $\mathbf{W}_{r}$ and $\mathbf{b}_{r}$ are described as weight and bias parameters respectively. The nonlinear activation functions are defined by $\boldsymbol{\Theta}_{r}$.\par
In the SLP-DNet design, the Lagrange multiplier associated with the equality constraint is wired across the network to provide additional flexibility. It is important to note that the architectures are the same for both non-robust and robust power minimization problems described in Sections \ref{non_robust_prob} and \ref{robust_problem} but differ in proximity barrier functions (PBFs). Therefore, to simplify our exposition, we build the structure of the learning framework based on (\ref{beam_update_relaxed}) and the DU framework described in \cite{bertocchi2020deep}, which gives rise to Algorithm \ref{algorithm_1}.
\begin{algorithm}
 \caption{Feed-forward-Backward Proximal IPM}
 \begin{algorithmic}[1]\label{algorithm_1}
 \renewcommand{\algorithmicrequire}{\textbf{Input:}}
 \renewcommand{\algorithmicensure}{\textbf{Output:}}
 \REQUIRE $\mathbf{w}_{1}^{[0]}$, ${\gamma}^{[0]}$, ${\lambda}^{[0]}$ and ${\mu}^{[0]}$
 \ENSURE  $\mathbf{w}_{1}$ 
 \\ \textit{Initialization} :
  \STATE randomly initialize $\mathbf{w}_{1}^{[0]}\in{\mathbb{R}^{2N_{t}\times1}}$, ${\mu}^{[0]}>0$, $\lambda^{[0]}>0$ and $\gamma^{[0]}>0$ $\forall\ {i}=1,\ \cdots,\ K$ 
  \FOR {$r=0$ to $L$}
  \STATE $\mathbf{w}_{1}^{[r+1]}=\text{prox}_{\gamma^{[r]}{\mu}^{[r]}B}\left(\mathbf{w}_{1}^{[r]}-\gamma^{[r]}\Delta{{D}(\mathbf{w}_{1}^{[r]},\ \lambda^{[r]})}\right).$
  \ENDFOR
 \RETURN $\mathbf{w}_{1}$ 
 \end{algorithmic} 
\end{algorithm}
\begin{algorithm}
 \caption{Proximity Barrier Operator of a Nonrobust SLP-DNet}
 \begin{algorithmic}[1]\label{algorithm_relaxed}
 \renewcommand{\algorithmicrequire}{\textbf{Input:}}
 \renewcommand{\algorithmicensure}{\textbf{Output:}}
 \REQUIRE $\mathbf{h}_{Ri}$, $\mathbf{h}_{Ii}$, $\Gamma_{i}$ and ${w}_{0}\ (\text{noise power})$
 \ENSURE  $\mathbf{w}_{1}$, $\gamma$, ${\mu}$ and $\lambda$
 \\ \textit{Initialization} :
  \STATE randomly initialize $\mathbf{w}_{0}\in{\mathbb{R}^{2N_{t}\times1}}$, ${\mu}^{[0]}>0$, $\lambda^{[0]}>0$ and $\gamma^{[0]}>0$ $\forall\ {i}=1,\ \cdots,\ K$.
  \STATE {Find the solution to (\ref{cubic_equation}) using Cardano formula}.
  \STATE \text{For every solution in step 2, compute its corresponding} \text{Barrier function using} \text{(\ref{prox_func_rel2})}.
  \STATE \text{Compute the Proximity Operator of the Barrier at} $\mathbf{w}_{0}$ \text{using (\ref{prox_op}), where  $    \Phi(\mathbf{w}_{1},\ \gamma,{\mu})=\text{prox}_{\gamma{\mu}{{B}}}{(\mathbf{w}_{1})}.$}
  \STATE \text{Compute the derivatives of the Proximity Operator} \text{w.r.t} $\mathbf{w}_{1}$, ${\mu}$ and $\gamma$ \text{using} (\ref{jacob_mat_relax}), (\ref{deriv_mu_rl}) and (\ref{deriv_gamma_rl}).
  \STATE {Update the training variables as follows:}
    \begin{enumerate}[label=(\alph*)]
      \item ${\mu}^{[r+1]}={\mu}^{[r]}-\eta\frac{\partial{\Phi}(\mathbf{w}_{1}^{[r]},\ \gamma^{[r]},\ {\mu}^{[r]})}{\partial{{\mu}^{[r]}}}$
      \item  $\gamma^{[r+1]}=\gamma^{[r]}-\eta\frac{\partial{\Phi}(\mathbf{w}_{1}^{[r]},\ \gamma^{[r]},\ {\mu}^{[r]})}{\partial{\gamma^{[r]}}}$
      \item  $\lambda^{[r+1]}=\lambda^{[r]}-\eta      \frac{\partial{{D}(\mathbf{w}_{1}^{[r]},\ \lambda^{[r]})}}{\partial{\lambda^{[r]}}}$ using (\ref{eq_constr_rl})
  \end{enumerate}
  \text{where} $\eta$ \text{is the learning rate}.
  \STATE \text{Use the results in step 6 and the Algorithm \ref{algorithm_1}} \text{to obtain the optimal precoding tensor}.
 \end{algorithmic} 
\end{algorithm}
As shown in Fig. \ref{fig:DNBF_Arch}, SLP-DNet has two main units; the parameter update module (PUM) and the auxiliary processing block (APB). The PUM has three core components associated with Lagrangian multipliers (equality and inequality constraints) and the training step-size, which are updated according to the following
\begin{equation}\label{update_func}
\mathcal{H}(\mathbf{w}_{1},{\mu},\gamma,\lambda)=\text{prox}_{\gamma^{[r]}{\mu}^{[r]}B}\left(\mathbf{w}_{1}^{[r]}-\gamma^{[r]}\Delta{{D}(\mathbf{w}_{1}^{[r]},\lambda^{[r]})}\right).
\end{equation}

Furthermore, the component that forms the barrier term is constructed with one convolutional layer, an average pooling layer, a fully connected layer, and a Softplus layer to curb the output to a positive real value to satisfy the inequality constraint. The APB unit is connected to the last \textit{r}-th block of the PUM in the form of a deep CNN to convert the output of the last parameter update block into a target transmit precoding vector. The APB architecture is made up of 3 convolution layers and 2 activation layers. In addition, a Batch Normalization layer is added between each convolutional layer and the activation layer to stabilize the mismatch in the distribution of the inputs caused by the internal covariate shift \cite{ioffe2015batch}. For every \textit{r} block (\textit{r}-th layer), there are three core components; $\mathcal{L}_{{\mu}}^{[r]}$, $\mathcal{L}_{\gamma}^{[r]}$ and $\mathcal{L}_{\lambda}^{[r]}$ associated with the learnable parameters (${\mu}$, $\gamma$ and $\lambda$), respectively as shown in Fig. \ref{fig:DNBF_Arch}. These components form a learning block for computing the barrier parameter (${\mu}$) associated with the inequality constraint, the step-size ($\gamma$) and the equality constraint ($\lambda$), if exists. To ensure that the constraints remain positive, a Softplus-sign function \cite{zheng2015improving}, $\text{Softplus}(z)=\ln{(1+\text{exp}(z))}$ is used.

The Softplus-sign function is a smooth approximation of the rectified linear unit (\textbf{ReLu}) activation function; and unlike the \textbf{ReLu} its gradient is never exactly equal to zero \cite{zheng2015improving}, which imposes an update on $\gamma$, ${\mu}$ and $\lambda$ during the backward propagation. The PBF for nonrobust SLP formulation is summarized in Algorithm \ref{algorithm_relaxed} below. A similar algorithm can also be adopted for a robust PBF using a robust formulation. Finally, the output from the auxiliary processing block is the precoding vector in the real domain. The relation: $\mathbf{w}_{1} = [\mathbf{w}_{R} \ -\mathbf{w}_{I}]^{T}$  is used to convert it to its equivalent complex domain for every SINR value of the \textit{i}-th user.
\subsubsection{Duality and Loss Function of the SLP Formulation}
In order to ease the formulation of the dual-problem of the original problem (\ref{P_relaxed1}), the left-hand-side of the inequality constraint is split into its equivalent positive and negative parts as follows 
\begin{equation}\label{P_relaxed4}
    \begin{aligned}
    & \underset{\mathbf{\{w_{1}\}}}{\text{min}}
    && {\norm{\mathbf{w}_{1}}_{2}^2} \\
    & \text{s.t.}
    && \boldsymbol{\Lambda}_{i}^{T}\boldsymbol{\Pi}\mathbf{w}_{1}\leq\left(\boldsymbol{\Lambda}_{i}^{T}\boldsymbol{\Pi}\mathbf{w}_{1}-\sqrt{\Gamma_{i}v_{0}}\right){\text{tan}{\phi}},\ \forall {i}\\
    &&& -\boldsymbol{\Lambda}_{i}^{T}\boldsymbol{\Pi}\mathbf{w}_{1}\leq\left(\boldsymbol{\Lambda}_{i}^{T}\boldsymbol{\Pi}\mathbf{w}_{1}-\sqrt{\Gamma_{i}v_{0}}\right){\text{tan}{\phi}},\ \forall {i}.
    \end{aligned}
\end{equation}
The Lagrangian of (\ref{P_relaxed4}) is defined as 
\begin{multline}\label{Lag_relaxed1}
\mathcal{L}_{\text{rl}}(\mathbf{w}_{1},\ \boldsymbol{\mu}_{1},\ \boldsymbol{\mu}_{2}) =\norm{\mathbf{w}_{1}}_{2}^{2} \\
+\boldsymbol{\mu}_{1}\left(\boldsymbol{\Lambda }^{T}_{i}\boldsymbol{\Pi }\mathbf{w}_{1}-\boldsymbol{\Lambda}_{i}^{T}\mathbf{w}_{1}\text{tan}{\phi}+\sqrt{\Gamma_{i}v_{0}}\right)\\
-\boldsymbol{\mu}_{2}\left(\boldsymbol{\Lambda }^{T}_{i}\boldsymbol{\Pi }\mathbf{w}_{1}+\boldsymbol{\Lambda}_{i}^{T}\mathbf{w}_{1}\text{tan}{\phi}-\sqrt{\Gamma_{i}v_{0}}\right),
\end{multline}
where $\boldsymbol{\mu}_{1}$ and $\boldsymbol{\mu}_{2}$ are the Lagrangian multipliers associated with the constraints and are related to the proximity barrier. The subscript \textit{`rl'} stands for relaxed phase rotation. It can be easily proven that the lower bound (\textbf{LB}) of (\ref{Lag_relaxed1}) is $\mathbcal{L}_\text{rl}(\mathbf{w}_1, \boldsymbol{\mu}_1, \boldsymbol{\mu}_2)\geq\boldsymbol{\mu}_{1}\Lambda_{i}\left(\Pi-\text{tan}{\phi}\right)-\boldsymbol{\mu}_{2}\Lambda_{i}\left(\Pi+\text{tan}{\phi}\right)$. From (\ref{Lag_relaxed1}), the optimal precoder is obtained by differentiating $\mathcal{L}_\text{rl}(\cdotp)$ w.r.t $\mathbf{w}_{1}$ and equating to zero. By doing so, the optimal precoder can be found as
\begin{equation}\label{optima_prec_rel}
\mathbf{w}_{1}=\frac{\left(\boldsymbol{\mu_{1}}^{T}+\boldsymbol{\mu_{2}}^{T}\right)\cdotp\boldsymbol{\Lambda}_{i}\text{tan}{\phi}-\left(\boldsymbol{\mu_{1}}^{T}-\boldsymbol{\mu_{2}}^{T}\right)\cdotp\boldsymbol{\Pi^{T}}\boldsymbol{\Lambda}_{i}}{2}.
\end{equation}

In the sequel, we show that (\ref{optima_prec_rel}) is used to generate the training input (precoding vector) by randomly initializing the Lagrangian multipliers ($\boldsymbol{\mu}_{1}$ and $\boldsymbol{\mu}_{2}$) and then train the network to learn their values that minimize the loss function (Lagrangian function). The loss function is modified by adding $\mathbcal{l}_{2}$-norm regularization over the weights to calibrate the learning coefficients in order to adjust the learning process. It should be noted that the regularization here is not aimed at addressing the problem of overfitting as in the case of supervised learning. However, regularization in an unsupervised learning is used to normalize and moderate weights attached to a neuron to help stabilize the learning algorithm \cite{wang2017regularization}. The loss function (\ref{Lag_relaxed1}) over $N$ training samples is thus expressed as
\begin{multline}\label{Lag_relaxed2}
\mathcal{L}_{\text{rl}}(\mathbf{w}_{1},\ \boldsymbol{\mu}_{1},\ \boldsymbol{\mu}_{2}) =\frac{1}{N}\sum^{N}_{i=1}\Vert \mathbf{w}_{1}\Vert_{2}^{2}\\
+\frac{1}{N}\sum^{N}_{i=1}\left(\boldsymbol{\mu}_{1}\left(\boldsymbol{\Lambda }^{T}_{i}\boldsymbol{\Pi }\mathbf{w}_{1}-\boldsymbol{\Lambda}_{i}^{T}\mathbf{w}_{1}\text{tan}{\phi}+\sqrt{\Gamma_{i}v_{0}}\right)\right)\\
-\frac{1}{N}\sum^{N}_{i=1}\left(\boldsymbol{\mu}_{2}\left(\boldsymbol{\Lambda }^{T}_{i}\boldsymbol{\Pi }\mathbf{w}_{1}+\boldsymbol{\Lambda}_{i}^{T}\mathbf{w}_{1}\text{tan}{\phi}-\sqrt{\Gamma_{i}v_{0}}\right)\right) \\
+\frac{\vartheta}{NL}\sum^{N}_{i=1}\sum_{i=1}^{L}\Vert \boldsymbol{\theta}_{i}\Vert_{2}^{2},
\end{multline}
where $\boldsymbol{\theta}_{i}$ are the trainable parameters of the \textit{i}-th layers associated with the weights and biases, and $\vartheta >0$ is the penalty parameter that controls the bias and variance of the trainable coefficients, $N$, $L$ is the number of training samples (batch size or the number of channel realization) and the number of layers, respectively. 
\subsection{Learning-Based SLP for Strict Angle Rotation}
In the previous subsection, we have presented SLP-DNet based on relaxed angle formulation. In this subsection, we provide a formulation for strict phase angle rotation where all the interfering signals align exactly to the phase the signal of interest (i.e. $\phi=0$ in Fig. \ref{fig:CI_GRAHICAL_REP}), the optimization problem is \cite{masouros2015exploiting}
\begin{equation} \label{p_st}
    \begin{aligned}
    & \underset{\mathbf{\{w_{1}\}}}{\text{min}}
    & & {\norm{\mathbf{w}_{1}}^2} \\
    & \text{s.t.}
    & & {\boldsymbol{\Lambda}_{i}^{T}\boldsymbol{\Pi}\mathbf{w}_{1}=0}\ ,\ \forall {i}\\
    &&& \boldsymbol{\Lambda}_{i}^{T}\mathbf{w}_{1}\geq \sqrt{\Gamma_{i}v_{0}} \ , \ \forall {i}.
    \end{aligned}
    \end{equation}
We observe that the inequality constraint in (\ref{p_st}) is affine. Based on this, the proximal barrier function for the strict phase rotation is
\begin{equation}\label{prox_func_st}
  B_\text{st}(\mathbf{w}_{1}) = \left \{
  \begin{aligned}
    &-\ln{\left(\boldsymbol{\Lambda}_{i}^{T}\mathbf{w}_{1}-\sqrt{{\Gamma}_{i} v_{0}}\right)}, && \text{if}\ \boldsymbol{\Lambda}_{i}^{T}\mathbf{w}_{1} \geq\sqrt{{\Gamma}_{i}v_{0}} \\
    &+\infty, && \text{otherwise.}
  \end{aligned} \right.
\end{equation}
The subscript \textit{`st'} represents strict phase rotation. Therefore, for every precoding vector $\mathbf{w}_{1} \in\mathbb{R}^{2N_{t}\times1}$, the proximity operator of ${\mu}\gamma{B}_\text{st}$ at $\mathbf{w}_{1}$ is given by
\begin{multline}\label{prox_op_strct}
    \Phi_\text{st}(\mathbf{w}_{1},{\mu},\gamma)=\mathbf{w}_{1}+ \\ \frac{\boldsymbol{\Lambda_{i}}^{T}\mathbf{w}_{1}-\sqrt{\Gamma_{i}v_{0}}-\sqrt{(\boldsymbol{\Lambda}_{i}^{T}\mathbf{w}_{1}-\sqrt{\Gamma_{i}v_{0}})^{2}+4\gamma{\mu}\norm{\boldsymbol{\Lambda}_{i}^{T}}_{2}^{2}}}{{{2\norm{\boldsymbol{\Lambda}_{i}}_{2}^{2}}}}\boldsymbol{\Lambda}_{i}.
\end{multline}
Similar to the steps in subsection \ref{Prox_Operator_relaxed}, the learning-based framework for SLP strict phase rotation is designed by finding the Jacobian matrix of $\Phi(\mathbf{w}_{1},{\mu},\gamma)$ with respect to $\mathbf{w}_{1}$, and the derivatives of $\Phi(\mathbf{w}_{1},{\mu},\gamma)$ with respect to $\gamma$ and $\mu$ can be easily obtained from (\ref{prox_op_strct}). The loss function over $N$ training batches is given by 
\begin{multline}\label{Loss_strict}
\mathcal{L}_\text{st}(\mathbf{w}_{1} ,\boldsymbol{\lambda} ,\ \boldsymbol{\mu} ) =\frac{1}{N}\sum^{N}_{i=1}\left(\Vert \mathbf{w}_{1}\Vert_{2} ^{2}+\boldsymbol{\lambda}\boldsymbol{\Lambda }^{T}_{i}\boldsymbol{\Pi}\mathbf{v}_{1}\right) + \\ \frac{1}{N}\sum^{N}_{i=1}\left(\boldsymbol{\mu}\left(\sqrt{\Gamma _{i} v_{0}} -\boldsymbol{\Lambda }^{T}_{i}\mathbf{w}_{1}\right)\right)+ 
\frac{\vartheta}{NL}\sum^{N}_{i=1}\sum_{i=1}^{L}\Vert \boldsymbol{\theta}_{i}\Vert_{2} ^{2},
\end{multline}
where $\boldsymbol{\mu}$ and $\boldsymbol{\lambda}$ are the Lagrangian multipliers for inequality and equality constraints, respectively. Finally, minimizing (\ref{Loss_strict}) with respect to $\mathbf{w}_{1}$ (differentiating $\mathcal{L}_\text{st}(\cdotp)$ w.r.t $\mathbf{w}_{1}$), gives the optimal precoder as
\begin{equation}\label{optima_prec_str}
\mathbf{w}_{1}=\frac{\boldsymbol{\mu}^{T}\cdotp\boldsymbol{\Lambda}_{i}-\boldsymbol{\lambda}^{T} \cdotp \boldsymbol{\Pi}\boldsymbol{\Lambda}_{i}}{2}.
\end{equation}
\section{Learning-Based Robust Power Minimization SLP with Channel Uncertainty}\label{robust_problem}\subsection{Channel Uncertainty and Channel error Model}
So far, we have derived the unsupervised learning scheme in which the uncertainty in estimating the channel coefficients is not considered. The exact CSI is often unobtainable in practice. To model the user's actual channel in the uncertainty region, we consider an ellipsoid $\xi$ such that
\begin{equation}\label{uncert_channel}
   \hat{\mathbf{h}}_{i}=\Bar{\mathbf{h}}_{i}+\Bar{\boldsymbol{e}}_{i} \ \forall{k},
\end{equation}
where $\Bar{\mathbf{h}}_{i}$ is the known CSI estimates at the BS and $\Bar{\boldsymbol{e}}_{i}$ denotes the channel error within the uncertainty region of the ellipsoid (i.e $ \hat{\mathbf{h}}_{i} \in \xi $). The model of the uncertainty ellipsoid with the center $\Bar{\mathbf{h}}_{i}$ is expressed as
\begin{equation}\label{uncert_ellipsoid}
   \xi =\left\{\Bar{\mathbf{h}}_{i} +\Bar{\mathbf{e}}_{i} |_{\| \Bar{\mathbf{e}}_{i}\leq 1 \| }\right\}.
\end{equation}
As shown in \cite{masouros2015exploiting}, the channel error is given by $\left\{\Bar{\mathbf{e}}_{i}: \norm{\Bar{\mathbf{e}}_{i}}_{2}^{2}\leq{\varsigma}^{2}_{i}\right\}$. It is important to note that the BS is assumed to have the knowledge about the channel error, excluding its corresponding error bound ${\varsigma}^{2}_{i}$. For details and formulation of the conventional robust BLP, we refer the reader to \cite{zheng2008robust}. 
\subsection{Robust Optimization-Based SLP Formulation}\label{robust_SLP}
The multi-cast constructive interference formulation of the power minimization problem for the worst-case CSI error is given by \cite{zheng2008robust}
\begin{equation} \label{robust_multi}
    \begin{aligned}
    & \underset{\mathbf{\{w\}}}{\text{min}}
    & & {\norm{\mathbf{w}}_{2}^2} \\
    & \text{s.t.}
    & & \Bigl|{\text{Im}\left(\hat{\mathbf{h}}_{i}^{T}\mathbf{w}\right)\Bigl|}-\left({\text{Re}\left(\hat{\mathbf{h}}_{i}^{T}\mathbf{w}\right)}-\sqrt{\Gamma_{i}v_{0}}\right)\text{tan}{\phi}\leq 0,\\
    &&& \forall \norm{\bar{\mathbf{e}}_{i}}^{2}\leq{\varsigma}_{i}^{2},\ \forall{i}.
    \end{aligned}
\end{equation}
The intractability of the constraint in (\ref{robust_multi}) can be effectively handled using convex optimization methods. Therefore, to guarantee that the robust constraint in (\ref{robust_multi}) is satisfied, it is modified as follows \cite{masouros2015exploiting}
\begin{equation}\label{robust_constraint}
\max _{\| \bar{e}_{i} \| ^{2} \leq \varsigma ^{2}_{i}}\left(\Bigl|\text{Im}\left(\hat{\mathbf{h}}^{T}\mathbf{w}\right)\Bigl| -\left(\text{Re}\left(\hat{\mathbf{h}}^{T}\mathbf{w}\right) -\sqrt{\Gamma v_{0}}\right)\text{tan} \phi \right) \leq 0.
\end{equation}

It is worth noting that the subscripts in (\ref{robust_constraint}) are ignored in order to simplify the problem formulation. By defining the equivalent real-valued channel and channel error vectors, the real and imaginary parts in the constraint can be decomposed into two separate constraints as explained in Section \ref{non_robust_prob} (see (\ref{Hw_Re}) and (\ref{Hw_Im})). Thus the robust formulation of the constraint is equivalent to two separate real-valued constraints as follows
\begin{equation}\label{robust_constraint1}
\boldsymbol{\Lambda}^{T}\mathbf{w}_{1}-\boldsymbol{\Lambda}^{T}\mathbf{w}_{2}\text{tan}{\phi}+\varsigma\norm{\mathbf{w}_{1}-\mathbf{w}_{2}\text{tan}{\phi}}_{2} + \sqrt{\Gamma v_{0}}\text{tan} \phi \leq 0,
\end{equation}
\begin{equation}\label{robust_constraint2}
-\boldsymbol{\Lambda}^{T}\mathbf{w}_{1}-\boldsymbol{\Lambda}^{T}\mathbf{w}_{2}\text{tan}{\phi}+\varsigma\norm{\mathbf{w}_{1}+\mathbf{w}_{2}\text{tan}{\phi}}_{2}+ \sqrt{\Gamma v_{0}}\text{tan} \phi \leq 0,
\end{equation}
where  $\boldsymbol{\Lambda}= \begin{bmatrix}
\bar{\mathbf{h}}_{R} & \bar{\mathbf{h}}_{I}
\end{bmatrix}^{T}$, $\mathbf{e}\overset{\Delta}{=}\begin{bmatrix}
\bar{\mathbf{e}}_{R} & \bar{\mathbf{e}}_{I}
\end{bmatrix}^{T}$ and $\hat{\mathbf{h}}=\bar{\mathbf{h}}_R+j\bar{\mathbf{h}}_I+\bar{\mathbf{e}}_R+j\bar{\mathbf{e}}_I$.
Finally, the robust CI formulation for power minimization problem becomes

\begin{equation} \label{robust_multi2}
    \begin{aligned}
    & \underset{\mathbf{\{w_{1},w_{2}\}}}{\text{min}}
    & & {\norm{\mathbf{w}_{1}}_{2}^2} \\
    & \text{s.t.}
    & & \text{Constraints}\ (\ref{robust_constraint1})\ \text{and}\  (\ref{robust_constraint2}), \ \forall{i}\\
    &&&  \text{where}\ \ \mathbf{w}_{1}=\boldsymbol{\Pi}\mathbf{w}_{2}.
    \end{aligned}
\end{equation}
\subsection{Proposed Unsupervised Learning-Based Robust SLP}  
In this subsection, we extend our proposed unsupervised learning formulation to a worst-case CSI-error to design a robust precoding scheme for the power minimization problem. As an extension of the previous formulations in subsection \ref{Prox_Operator_relaxed}, the focus here is to derive a PBF for the robust learning-based precoding scheme. Substituting for $\mathbf{w}_{1}$ in (\ref{robust_multi2}), we have
\begin{multline}\label{robust_constraint12}
\left(\boldsymbol{\Lambda}^{T}\boldsymbol{\Pi}-\boldsymbol{\Lambda}^{T}\text{tan}{\phi}\right)\mathbf{w}_{2}+\varsigma\norm{\left(\boldsymbol{\Pi}-\text{tan}{\phi}\right)\mathbf{w}_{2}}_{2}\\
+ \sqrt{\Gamma v_{0}}\text{tan} \phi \leq 0,
\end{multline}
\begin{multline}\label{robust_constraint22}
-\left(\boldsymbol{\Lambda}^{T}\boldsymbol{\Pi}+\boldsymbol{\Lambda}^{T}\text{tan}{\phi}\right)\mathbf{w}_{2}+\varsigma\norm{\left(\boldsymbol{\Pi}+\text{tan}{\phi}\right)\mathbf{w}_{2}}_{2}\\ +\sqrt{\Gamma v_{0}}\text{tan} \phi \leq 0.
\end{multline}
Apparently, the constraints (\ref{robust_constraint12}) and (\ref{robust_constraint22}) are bounded by the $\mathcal{l}_{2}$-norm. Therefore, problem (\ref{robust_multi2}) is rewritten as
\begin{equation} \label{robust_multi3}
    \begin{aligned}
    & \underset{\mathbf{\{w_{2}\}}}{\text{min}}
    & & {\norm{\mathbf{w}_{2}}_{2}^2} \\
    & \text{s.t.}
    & & \text{Constraints}\ (\ref{robust_constraint12})\ \text{and}\  (\ref{robust_constraint22}), \ \forall{i}.
    \end{aligned}
\end{equation}
The resulting barrier function of the corresponding constraints of (\ref{robust_multi3}) is the sum of the individual barrier functions associated with the two inequality constraints. We begin by introducing the feasible set of solutions bounded by the Euclidean ball.

\subsubsection{Bounded Euclidean ball Constraint}
Suppose a problem whose set of feasible solutions is bounded by the Euclidean ball \cite{parikh2014proximal}
\begin{equation}\label{norm_constraint}
  \mathbcal{C}=\{\mathbf{z}\in\mathbb{R}^{n}\big|\norm{\mathbf{z}-\mathbf{x}}_{2}\leq\beta\},  
\end{equation}
where $\beta>0$ and $\mathbf{x}\in\mathbb{R}^{n}$.
Let $\gamma>0$ and $\mu>0$ be the step-size and the Lagrange multiplier associated with the inequality constraint, respectively. Then the barrier function is expressed as \cite{parikh2014proximal}
\begin{equation}\label{prox_func_robust}
  {B(\mathbf{z})} = \left \{
  \begin{aligned}
    &-\ln{\left(\beta-\norm{\mathbf{z}-\mathbf{x}}_{2}\right)}, && \text{if}\ \norm{\mathbf{z}-\mathbf{x}}_{2}<\beta, \\
    &+\infty, && \text{otherwise}
  \end{aligned} \right.
\end{equation}

For simplicity, we let $\mathbf{Q}_{1}=\left(\boldsymbol{\Pi}-\mathbf{I}_{2N_{t}}\text{tan}{\phi}\right)$ and $\mathbf{Q}_{2}=\left(\boldsymbol{\Pi}+\mathbf{I}_{2N_{t}}\text{tan}{\phi}\right)$. Based on (\ref{prox_func_robust}), the barrier function corresponding to the constraint (\ref{robust_constraint12}) is written at the bottom of the page.
\begin{figure*}[!b]
\hrule
\vspace{1mm}
\begin{equation}\label{prox_func_robust1}
  {B_{1}(\mathbf{w}_{2})} = \left \{
  \begin{aligned}
    &-\ln{\left(-\sqrt{\Gamma v_{0}}\text{tan}{\phi}-\left(\boldsymbol{\Lambda}^{T}\mathbf{Q}_{1}\mathbf{w}_{2}+\varsigma\norm{\mathbf{Q}_{1}\mathbf{w}_{2}}_{2}\right)\right)},&& \text{if}\ \boldsymbol{\Lambda}^{T}\mathbf{Q}_{1}\mathbf{w}_{2}+\varsigma\norm{\mathbf{Q}_{1}\mathbf{w}_{2}}_{2}<-\sqrt{\Gamma v_{0}}\text{tan}{\phi} \\
    &+\infty && \text{otherwise}
  \end{aligned} \right.
\end{equation}
\end{figure*}
In the case of constraint (\ref{robust_constraint22}), similar expression is also written for ${B_{2}(\mathbf{w}_{2})}$ using $\mathbf{Q}_{2}$. 
The resulting barrier function is thus 
\begin{equation}\label{total_b_func}
    {B}_{\text{robust}}(\mathbf{w}_{2})={B_{1}(\mathbf{w}_{2})}+{B_{2}(\mathbf{w}_{2})}
\end{equation}
Without loss of generality, the constraints (\ref{robust_constraint12}) and (\ref{robust_constraint22}) can be further written as
\begin{equation}\label{c1}
    \boldsymbol{\Lambda}^{T}\mathbf{Q}_{1}\mathbf{w}_{2}+\varsigma\norm{\mathbf{Q}_{1}\mathbf{w}_{2}}_{2}+\sqrt{\Gamma v_{0}}\text{tan}{\phi}\leq0,
\end{equation}
\begin{equation}\label{c2}
    \boldsymbol{\Lambda}^{T}\mathbf{Q}_{2}\mathbf{w}_{2}+\varsigma\norm{\mathbf{Q}_{2}\mathbf{w}_{2}}_{2}+\sqrt{\Gamma v_{0}}\text{tan}{\phi}\leq0.
\end{equation}
It can be seen that the upper bound of the two constraints (\ref{c1}) and (\ref{c2}) is zero, Therefore, the effective proximity operator of (\ref{total_b_func}) is obtained the by squaring (\ref{c1}) and (\ref{c2}) and adding the results. Following similar steps presented in subsection \ref{Prox_Operator_relaxed}, we obtain the proximity operator of the barrier for the robust SLP-DNet (see Appendix \ref{appendix} for details).   

\subsubsection{Loss Function of the Robust Power Minimization Problem}
The training loss function is the Lagrangian of (\ref{robust_multi3}), and can be written as
\begin{equation} \label{robust_multi4}
    \scalebox{.94}{$\begin{aligned}
    & \underset{\mathbf{\{w_{2}\}}}{\text{min}}
    & & {\norm{\mathbf{w}_{2}}_{2}^2} \\
    & \text{s.t.}
    & & \boldsymbol{\Lambda}^{T}\mathbf{Q}_{1}\mathbf{w}_{2}+\varsigma\norm{\mathbf{Q}_{1}\mathbf{w}_{2}}_{2}+\sqrt{\Gamma v_{0}}\text{tan}{\phi}\leq0\  \forall{i}\\ 
    &&& \boldsymbol{\Lambda}^{T}\mathbf{Q}_{2}\mathbf{w}_{2}+\varsigma\norm{\mathbf{Q}_{2}\mathbf{w}_{2}}_{2}+\sqrt{\Gamma v_{0}}\text{tan}{\phi}\leq0\ 
    \forall{i}.
    \end{aligned}$}
\end{equation}
Therefore, the loss function of (\ref{robust_multi4}) is the regularized Lagrangian parameterized by $\boldsymbol{\theta}_{i}$ over the entire layers
\begin{multline}\label{Lag_robust_reg}
\mathcal{L}_{\text{robust}}(\mathbf{w}_{2},\ \boldsymbol{\mu}_{1},\ \boldsymbol{\mu}_{2}) =\frac{1}{N}\sum^{N}_{i=1}\Vert \mathbf{w}_{2}\Vert_{2}^{2} \\
+\frac{\boldsymbol{\mu}_{1}}{N}\sum^{N}_{i=1}\left({\varsigma}^{2}\norm{\mathbf{Q}_{1}\mathbf{w}_{2}}_{2}^{2}-\left(\sqrt{\Gamma v_{0}}\text{tan}{\phi}-\boldsymbol{\Lambda}^{T}\mathbf{Q}_{1}\mathbf{w}_{2}\right)^{2}\right)\\
+\frac{\boldsymbol{\mu}_{2}}{N}\sum^{N}_{i=1}\left({\varsigma}^{2}\norm{\mathbf{Q}_{2}\mathbf{w}_{2}}_{2}^{2}-\left(\sqrt{\Gamma v_{0}}\text{tan}{\phi}-\boldsymbol{\Lambda}^{T}\mathbf{Q}_{2}\mathbf{w}_{2}\right)^{2}\right)\\
+\frac{\vartheta}{NL}\sum^{N}_{i=1} \sum_{i=1}^{L}\Vert \boldsymbol{\theta}_{i}\Vert_{2}^{2}.
\end{multline}
The minimum of (\ref{Lag_robust_reg}) with respect to $\mathbf{w}_{2}$ by equating its derivative to zero
\begin{multline}\label{Lag_robust2}
    \left(1+\left(\boldsymbol{\mu}_{1}\norm{\mathbf{Q}_{1}}_{2}^{2}+\boldsymbol{\mu}_{2}\norm{\mathbf{Q}_{2}}_{2}^{2}\right)\left({\varsigma}^{2}-\boldsymbol{\Lambda}^{T}\boldsymbol{\Lambda}\right)\right){\mathbf{w}_{2}}=\\
    -\left(\boldsymbol{\mu}_{1}\mathbf{Q}_{1}+\boldsymbol{\mu}_{2}\mathbf{Q}_{2}\right)\boldsymbol{\Lambda}\sqrt{\Gamma{v}_{0}\text{tan}{\phi}}.
\end{multline}
For convenience, we redefine the real matrices and vectors as 
$\begin{bmatrix}\| \mathbf{Q}_{1} \|_{2}^{2} & 
 \| \mathbf{Q}_{2} \|_{2}^{2}\end{bmatrix} =\bar{\mathbf{q}}_{\text{norm}}$; $\begin{bmatrix}\mathbf{Q}_{1} & \mathbf{Q}_{2} \end{bmatrix} =\bar{\mathbf{Q}}$ and $\begin{bmatrix}\boldsymbol{\mu}_{1} & \boldsymbol{\mu}_{2} \end{bmatrix} =\bar{\boldsymbol{\mu}}$.
With these new notations, (\ref{Lag_robust2}) is simplified to 
\begin{equation}\label{Lag_robust3}
    \left(\mathbf{I}_{2N_{t}}+\mathbf{\bar{q}}_{\text{norm}}\boldsymbol{\bar{\mu}}^{T}\left({\varsigma}^{2}-\boldsymbol{\Lambda^{T}\Lambda}\right)\right){\mathbf{w}_{2}}=-\boldsymbol{\Lambda}\mathbf{\bar{Q}}{\boldsymbol{\bar{\mu}}}^{T}\sqrt{\Gamma{w}_{0}}{\text{tan}{\phi}}
\end{equation}
From (\ref{Lag_robust3}), the optimal transmit precoder is thus
\begin{equation}\label{robust_optimal}  
    \mathbf{w}_{2}=-\boldsymbol{\Lambda}\mathbf{\bar{Q}}{\boldsymbol{\bar{\mu}}}^{T}{\mathbf{A}}^{-1}\sqrt{\Gamma{v}_{0}}{\text{tan}{\phi}},
\end{equation}
where ${\mathbf{A}}=\left(\mathbf{I}_{2N_{t}}+\mathbf{\bar{q}}_{\text{norm}}\boldsymbol{\bar{\mu}}^{T}\left({\varsigma}^{2}-\boldsymbol{\Lambda^{T}\Lambda}\right)\right)$. Note that the Lagrange multipliers $\boldsymbol{\mu}_{1}$ and $\boldsymbol{\mu}_{2}$ are associated with the barrier term and are randomly initialized from a uniform distribution.

\section{Data Generation, Training and Computational Complexity}\label{data_train_complexity}
\subsection{ Dataset Generation}
\begin{figure}[!tb]
    \centering
    \includegraphics[width=3.56in,height=1.7in]{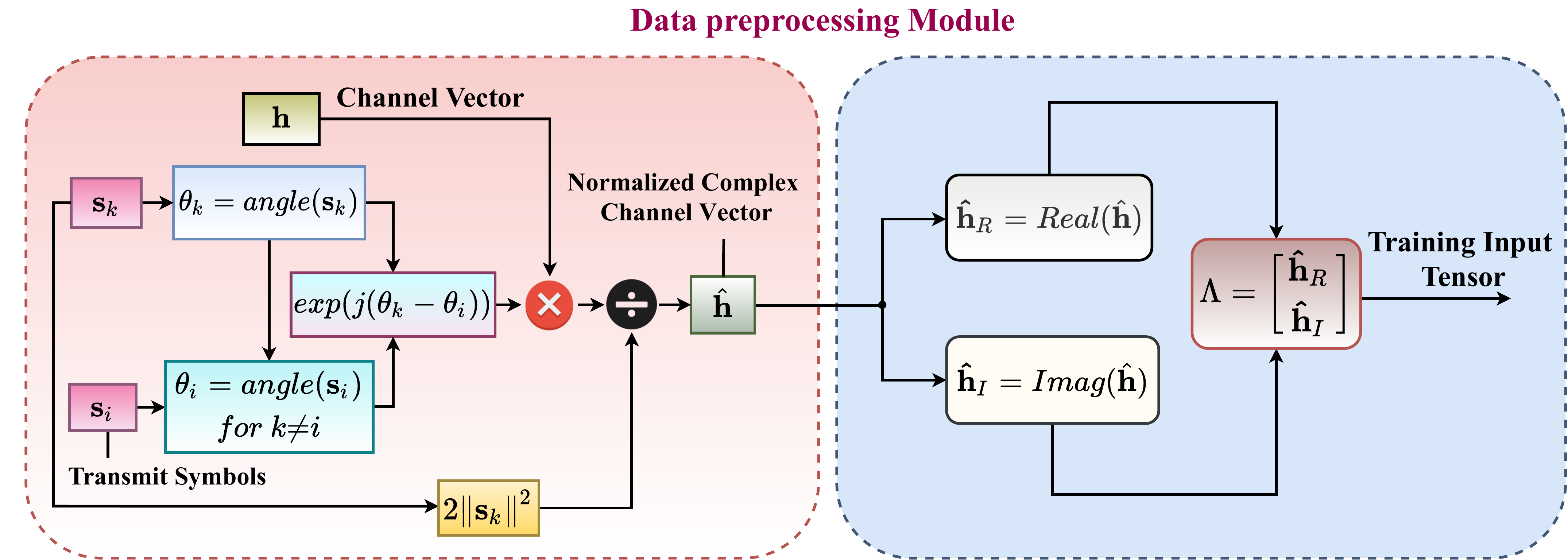}
    \caption{Dataset Generating Block.}
    \label{fig:DATASET_Block}
\end{figure}

The channel coefficients are used to form a dataset and are generated randomly from a normal distribution with zero mean and unit variance. The data input tensor is obtained using (\ref{h_hat}). We summarize the entire dataset preprocessing procedure in Fig. \ref{fig:DATASET_Block}. It can be observed that the input dataset is normalized by the transmit data symbol so that data entries are within the nominal range, and this could potentially aid the training.

\subsection{SLP-DNet Training and Testing}\label{training}
The training of DNNs generally involves three steps: forward propagation, backward propagation, and parameter update \cite{shalev2014understanding}. Except where necessary, the training SINR is drawn from a random uniform distribution to enable learning over a wide range of SINR values. The PUM contains \textit{r} blocks, which form a learning layer. Therefore, each block contains three core components and is trained block-wise for \textit{l} number of iterations.\par
Similarly, the APB is trained for \textit{k} iterations. It is important to note that the number of training iterations of the parameter update module may not necessarily be equal to that of the APB. We train the PUM for 15 iterations and the APB for 10 iterations. To improve the training efficiency, we modify the learning rate by a factor $\alpha \in \mathbb{R}^{+}$ for every training step. All the training is done with a stochastic gradient descent algorithm using Adam optimizer \cite{shalev2014understanding}. Since the learning is done in an unsupervised fashion, the loss function is the Lagrangian function's statistical mean over the entire training batch samples. During the inference, a feed-forward pass is performed over the entire architecture using the learned Lagrangian multipliers to calculate the precoding vector using (\ref{optima_prec_rel}) and (\ref{robust_optimal}) for both SLP and robust SLP formulations, respectively. Finally, at inference, the trained model is run with different SINR values to obtain the required optimal precoding matrix. 

\subsection{Computational Complexity Evaluation} 
In this subsection, we analyze and compare the computational costs of the conventional BLP, optimization-based SLP, and the proposed SLP-DNet schemes. The complexities are evaluated in terms of the number of real arithmetic operations involved. For ease of analysis, we convert the SOCP (\ref{P_relaxed1}) into a standard linear programming (LP)
\begin{equation} \label{P_complexity}
    \begin{aligned}
    & \underset{\mathbf{\{z\}}}{\text{min}}
    & & \mathbf{c}^{T}\mathbf{z} \\
    & \text{s.t.}
    & & \mathbf{c}_{k}^{T}\mathbf{z}\leq{-\text{tan}\phi\sqrt{\Gamma_{i}v_{0}}}\ ,\ \forall {i}
    \end{aligned}
\end{equation}
where $\displaystyle \mathbf{c=\begin{bmatrix}
0 & \mathbf{w}_{1}^{T}
\end{bmatrix}}^{T}\in \mathbb{R}^{(2N_{t}+1)\times{1}}$, $\mathbf{z} =\begin{bmatrix}
1 & \mathbf{w}_{1}
\end{bmatrix}^{T}\in \mathbb{R}^{(2N_{t}+1)\times1}$, $\mathbf{c}_{k}=\left[
\left|\boldsymbol{\Lambda }_{i}^{T}\boldsymbol{\Pi }\mathbf{w}_{1}\right|\ \boldsymbol{\Lambda}_{i}^{T}\tan \phi 
\right]^{T}\in \mathbb{R}^{(2N_{t}+1)\times1}$ and $\mathbf{W}=[\mathbf{w}_{11}, \cdots,\mathbf{w}_{1K}]; \ \ \forall{i=1,\cdots,K}$. 
The complexity per iteration for solving convex optimization via IPM is dominated by the formation ($\mathtt{C}_\text{form}$) and factorization ($\mathtt{C}_\text{fact}$) of the matrix coefficients of $m$ linear equations in $m$ unknowns \cite{wang2014outage}. For generic IPMs, the complexity is expressed as \cite{wang2014outage}
\begin{equation}\label{cost1}
    \mathtt{C}_\text{total}=\mathtt{C}_\text{iter}\cdot\left(\mathtt{C}_\text{form}+\mathtt{C}_\text{fact}\right)
\end{equation}
where $\mathtt{C}_\text{iter}$ is the iteration complexity required to attain an optimal solution. For a given optimal target accuracy, $\epsilon>0$, $\mathtt{C}_\text{iter}$ is given by
\begin{equation}\label{cost_iter}
    \mathtt{C}_\text{iter}=\sqrt{\sum_{j=1}^{N_\text{lc}}{d}_{j}+2N_\text{sc}}\times{\ln\left(\frac{1}{\epsilon}\right)}
\end{equation}
where $d$ is the dimension of the constraints, $N_\text{lc}$ and $N_\text{sc}$ are the numbers of linear inequality matrix and second order cone (SOC) constraints, respectively. The costs of formation and factorization of matrix are respectively given by \cite{wang2014outage}
\begin{equation}\label{cost_form_fact}
    \mathtt{C}_\text{form}=\underbrace{m\sum_{j=1}^{N_\text{lc}}d_{j}^{3}+m^{2}\sum_{j=1}^{N_\text{lc}}d_{j}^{2}}_\text{due to $N_\text{lc}$} +\underbrace{m\sum_{j=1}^{N_\text{sc}}d_{j=1}^{2}}_\text{due to $N_\text{sc}$}; \ \mathtt{C}_\text{fact}=m^{3}.
\end{equation}
Specifically, we observe that problem (\ref{P_complexity}) has $K$ constraints with dimension $2N_{t}+1$. Therefore, using (\ref{cost_iter}) and (\ref{cost_form_fact}), the total computational complexity is thus
\begin{equation}
    \scalebox{0.96}{$\mathtt{C}_\text{total}=\sqrt{2N_{t}+1}\left[m(2N_{t}+1)+m(2N_{t}+1)^{2}+m^{3}\right]\ln{\left(\frac{1}{\epsilon}\right)}$.}
\end{equation}
The complexity of BLP can be derived in a similar way and is shown directly in Table \ref{tab:complexity-table}.
Conversely, the complexity of the proposed SLP-DNet schemes is the sum of PUM and the APB complexities. Moreover, the complexity of the PUM is dominated by the costs of computing the \textit{`log barrier'} and the feed-forward pass of the shallow CNN (see Table \ref{tab:proximity_barrier_NN}) that makes up the barrier term associated with the inequality constraint. Similarly, the complexity of the APB is also obtained by computing the arithmetic operations involved during the forward pass of the deep CNN (see Table \ref{tab:auxiliary_NN}). To derive the analytical complexity of SLP-DNet, we assume a sliding window is used to perform the dominant computation of the convolution operation in the CNN and ignore the nonlinear computational overhead due to activations. Therefore, the total computational complexity is expressed as
\begin{multline}
C_{SLP-DNet}=C_{\text{log-br}}+\\2\sum^{L_\text{conv}}_{l=1}n_\text{h}^{[l-1]}n_\text{w}^{[l-1]}\left[C_\text{in}^{[l-1]}f^{[l]2}+1\right]C_\text{out}^{[l]}+\\\sum^{L_\text{fc}}_{j=1}\left(2M_\text{in}^{[j-1]}+1\right)M_\text{out}^{[i]}
\end{multline}
where $n_\text{h}$, $n_\text{w}$, $f$, $C_\text{in}$ and $C_\text{out}$ are the height, width of the input tensor, kernel size, number of input and output channels, respectively. Similarly, $L_\text{conv}$, $L_\text{fc}$, $M_\text{in}$ and $M_\text{out}$ are the number of convolution and fully connected (FC) layers, number of input and output neurons in the FC layer, respectively. $C_{\text{log-br}}$ denotes the complexity of the \textit{`log-barrier'} function. Table \ref{tab:complexity-table} shows the summary of the computational complexities of our proposals and the benchmark precoding schemes. As an illustration, we consider the case of a symmetrical system $(N_{t}=K=n)$, and show that the proposed approach has substantially reduced computational complexity of $\mathbcal{O}(n^{3})$, while the optimization-based SLP approach of $\mathbcal{O}(n^{6.5})$ and the conventional BLP is $\mathbcal{O}(n^{7.5})$. 
\begin{table*}[hbt!]
\resizebox{0.95\textwidth}{!}{\begin{minipage}{\textwidth}
\caption{Complexity analysis of proposed SLP-DNet and benchmark SLP schemes.}
\label{tab:complexity-table}
\centering
\begin{tabular}{l|l|l}
    \hline
    Problem  & Arithmetic Operations (term; $m=\mathcal{O}(2N_{t}K)$) & Complexity Order ($n=N_{t}=K$)\\
    \hline
    \hline
     Conventional BLP & $\sqrt{(4N_{t}+K+2)}\left[m(2N_{t}+1)+m(2N_{t}+1)^{2}+m(K+1)^{2}+m^{3}\right]\ln{\left(\frac{1}{\epsilon}\right)}$ & $\mathcal{O}(n^{6.5})$\\
    \hline
    SLP Optimization-based  &   $\sqrt{2N_{t}+1}\left[m(2N_{t}+1)+m(2N_{t}+1)^{2}+m^{3}\right]\ln{\left(\frac{1}{\epsilon}\right)}$ & $\mathcal{O}(n^{6.5})$\\
    \hline
    SLP-DNet &  $4K^{2}N_{t}+42K^{2}+48KN_{t}+512K+2$ & $\mathcal{O}(n^{3})$\\
    \hline
    SLP-DNet Strict & $4K^{2}N_{t}+39K^{2}+46KN_{t}+512K+2$ & $\mathcal{O}(n^{3})$ \\
    \hline
     Robust Conventional BLP &  $\sqrt{2K(2N_{t}+1)}\left[mK(2N_{t}+1)^{3}+m^{2}K(2N_{t}+1)^{2}+m^{3}\right]\ln{\left(\frac{1}{\epsilon}\right)}$ & $\mathcal{O}(n^{7.5})$\\
     \hline
    Robust SLP Optimization-based    &  $\sqrt{2(2N_{t}+1)}\left[2mK(2N_{t}+1)^{2}+m^{3}\right]\ln{\left(\frac{1}{\epsilon}\right)}$ & $\mathcal{O}(n^{6.5})$\\
    \hline
    Robust SLP-DNet & 
    $16KN_{t}^{2}+42K^{2}+48KN_{t}+512K$ & $\mathcal{O}(n^{3})$\\
    \hline
\end{tabular}
\end{minipage}}
\end{table*}

\section{Simulation, Results and Discussion}\label{results}
\subsection{Simulation Set-up}
We consider a single-cell MISO downlink in which the BS is equipped with four antennas ($N_{t}=4$) that serve $K=4$ single users. We generate 50,000 training and 2000 test samples of Rayleigh fading channel coefficients, respectively drawn from the same statistical distribution. The transmit data symbols are modulated using QPSK and 8PSK modulation schemes. The training SINR is randomly drawn from uniform distribution $\Gamma_{\text{train}} \sim \mathcal{U}(\Gamma_\text{low},\,\Gamma_\text{high})$. Adam optimizer \cite{shalev2014understanding} is used for stochastic gradient descent algorithm with Lagrangian function as a loss metric.\par 
Furthermore, a parametric rectified linear unit (\textbf{PReLu}) activation function is used for both convolutional and fully connected layers instead of the traditional \textbf{ReLu} function. The reason for this is to address the problem of dying gradient \cite{shalev2014understanding}. The learning rate is reduced by a factor $\alpha=0.65$ after every iteration to aid the learning algorithm to converge faster. The learning models are implemented in Pytorch 1.7.1 and Python 3.7.8 on a computer with the following specifications: Intel(R) Core (TM) i7-6700 CPU Core, 32.0GB of RAM. Table \ref{tab:experimental_parameter} summarizes the simulation parameters, while Tables \ref{tab:proximity_barrier_NN} and \ref{tab:auxiliary_NN} depict the NN component settings of the SLP-DNet.

\begin{table}[!htb]
\renewcommand{\arraystretch}{1.3}
\caption{Simulation settings}
\label{tab:experimental_parameter}
\centering
\begin{tabular}{l|l}
    \hline
    Parameters  &  Values\\
    \hline
    \hline
    Training Samples    &  50000 \\
    \hline
    Batch Size (B)   &  200 \\
    \hline
    Test Samples    &  2000 \\
    \hline
    Training SINR range    &  0.0dB - 45.0dB \\
    \hline
    Test SINR range (\textit{i}-th user SINR)    &  0.0dB - 35.0dB\\ 
    \hline
    Optimizer    &  SGD with Adam \\
    \hline
    Initial Learning Rate $\eta$   &  0.001 \\
    \hline
    Learning Rate decay factor $\alpha$  &  0.65 \\
    \hline
    Weight Initializer    &  Xavier Initializer \\
    \hline
    Number of blocks in the parameter \\ 
    update unit & $B_{r}=2$ \\
    \hline
    Training Iterations for each block \\
    of the parameter update unit   &  15 \\
    \hline
    Training iterations for the auxiliary unit &  10 \\
    \hline
\end{tabular}
\end{table}
\begin{table}[!htb]
\renewcommand{\arraystretch}{1.3}
\caption{Proximity Barrier Function NN Layout}
\label{tab:proximity_barrier_NN}
\centering
\begin{tabular}{l|l}
    \hline
    Layer  &  Parameter, $\text{kernel size}=3\times3$\\
    \hline
    \hline
    Input Layer    &  Input size $(\text{B},\ 1,\ 2N_{t},\ K)$ \\
    \hline
    Layer 1: Convolutional    & Size $(\text{B},1,K,20)$; zero padding\\
    \hline
    Layer 2: Average Pooling    & Size $((1,\ 1),\ \text{stride}=(1,\ 1))$\\ 
    \hline
    Layer 3: Activation    & Soft-Plus \\ 
    \hline
    Layer 4: Flat
    &  Size $(\text{B}\times20\times K^{2})$ \\
    \hline
    Layer : Fully-connected  & Size$(\text{B}\times20\times K^{2},\ 1)$\\
    \hline
    Layer 5: Activation    & Soft-Plus function\\ 
    \hline
\end{tabular}
\end{table}

\begin{table}[!htb]
\renewcommand{\arraystretch}{1.3}
\caption{Auxiliary Processing Block (APB) NN Structure}
\label{tab:auxiliary_NN}
\centering
\begin{tabular}{l|l}
    \hline
    Layer  &  Parameter, $\text{kernel size}=3\times3$\\
    \hline
    \hline
    Input Layer    &  Input size $(\text{B},\ 1,\ 2N_{t},\ K)$ \\
    \hline
    Layer 1: Convolutional    & Size $(\text{B},\ 1,\ K,\ 64)$,\\ & $\text{dilation} = 1$ and unit padding\\
    \hline
    Layer 2: Batch Normalization
    &  $\text{eps}=10^{-6}$, $\text{momentum}=0.1$\\
    \hline
    Layer 3: Activation   &  PReLu \\
    \hline
    Layer 4: Convolutional    & Size $(\text{B},\ 1,\ 64,\ 2N_{t}K)$,\\ &
    $\text{dilation} = 1$ and unit padding\\
    \hline
    Layer 5: Batch Normalization
    &  $\text{eps}=10^{-6}$, $\text{momentum}=0.1$\\
    \hline
    Layer 6: Activation   &  PReLu \\
    \hline
    Layer 7: Convolutional    & Size $(\text{B},\ 1,\ 2N_{t}K,\ 1)$,\\ & $\text{dilation} = 1$ and unit padding\\
    \hline
\end{tabular}
\end{table}

\subsection{Performance Evaluation of Non-Robust SLP-DNet}\label{nonrobust_performance}
In this subsection, we evaluate the performance of our proposed unsupervised learning framework for nonrobust scenario against the benchmark algorithms \cite{bjornson2014optimal, masouros2015exploiting, zheng2008robust} for both strict and relaxed angle rotations.\par
\begin{figure}[!t]
\centering
  \includegraphics[width=3.25in,height=2.5in]{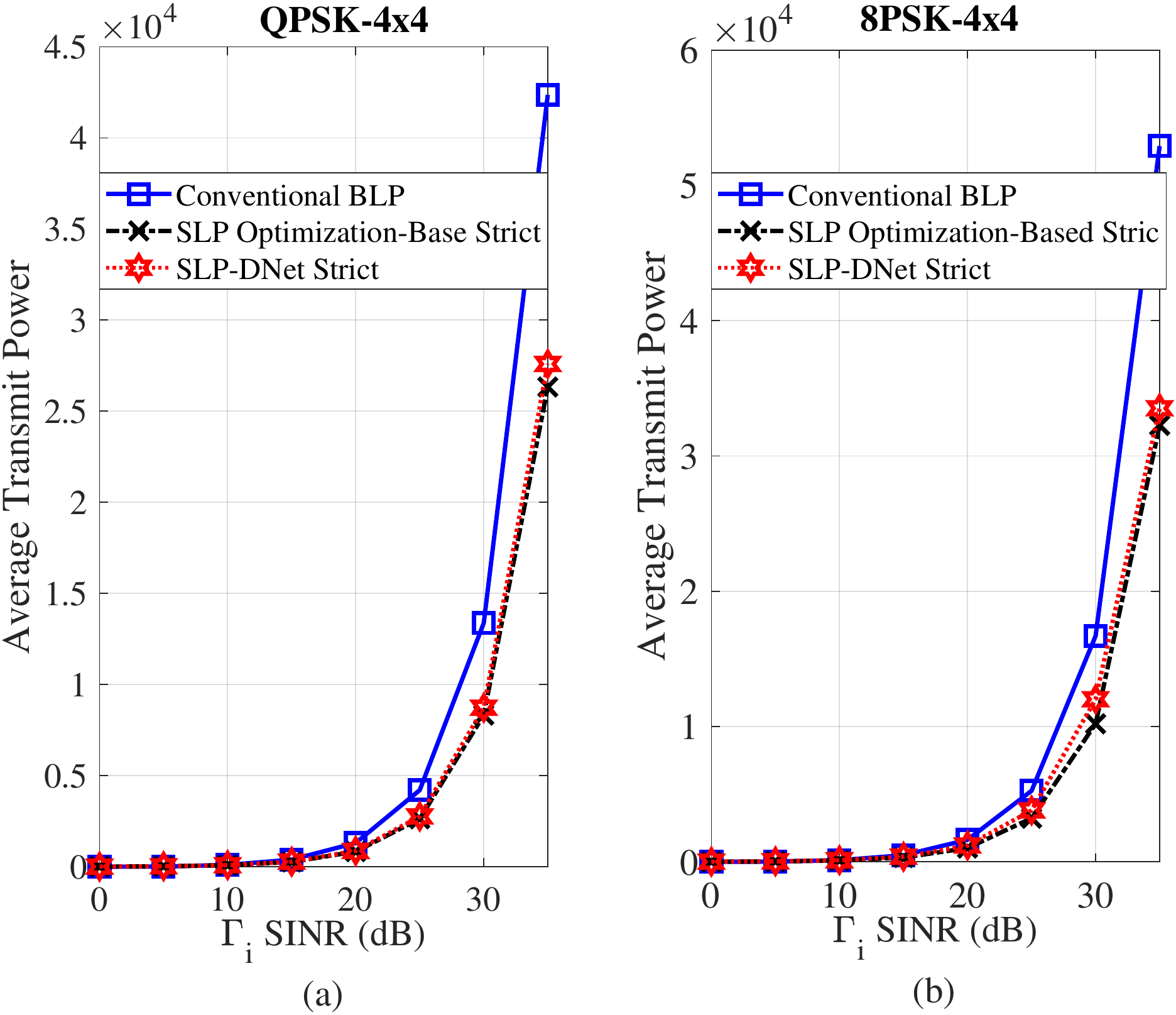}
\caption{Transmit Power vs SINR averaged over 2000 test samples for conventional BLP, SLP optimization-based and nonrobust SLP-DNet schemes for $\mathbb{M}$-PSK modulation with $N_{t}=4$, $K=4$ under strict angle rotation.}
\label{fig:POWER_vs_SINR_STRICT}
\end{figure}

Firstly, we compare the average transmit power of the conventional BLP (\ref{conv_power_min}), the SLP optimization-based problems (\ref{CI_optimization}), (\ref{P_relaxed1}) and the SLP-DNet precoding scheme based on (\ref{beam_update_relaxed}) and Algorithm \ref{algorithm_relaxed}. The performances of SLP-DNet and the benchmark schemes (conventional BLP and SLP optimization-based) for strict angle rotation are shown in Fig. \ref{fig:POWER_vs_SINR_STRICT}. It can be observed that the transmit power of the proposed SLP-DNet closely matches the optimisation based SLP, both with significant gains against BLP.\par
\begin{figure}[!t]
\centering
  \includegraphics[width=3.25in,height=2.5in]{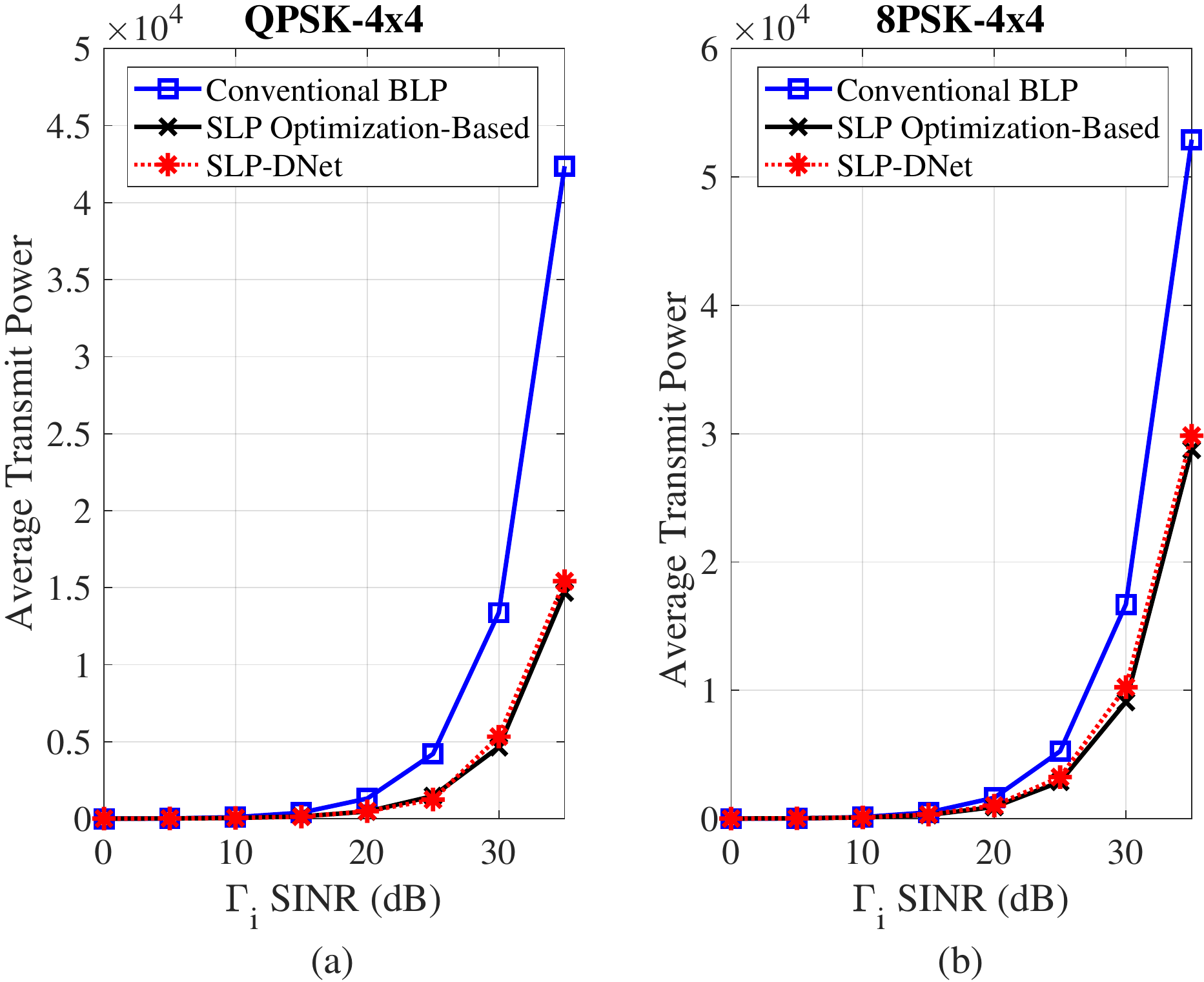}
\caption{Transmit Power vs SINR averaged over 2000 test samples for conventional BLP, SLP optimization-based and nonrobust SLP-DNet schemes for $\mathbb{M}$-PSK modulation with $N_{t}=4$, $K=4$ under relaxed angle rotation.} 
\label{fig:POWER_vs_SINR_RELAXED}
\end{figure}

 
Similarly, we discern the same trend in Fig. \ref{fig:POWER_vs_SINR_RELAXED} for the relaxed angle scenario as observed in Fig. \ref{fig:POWER_vs_SINR_STRICT}. Accordingly, we find from Fig. \ref{fig:POWER_vs_SINR_RELAXED} that the relaxed angle formulation offers significant power savings over the strict angle formulation and is therefore adopted in the subsequent experiments. Furthermore, at 30dB, the performance of SLP-DNet is within 5\% of the SLP optimization-based solution. Thus, while the SLP optimization-based offers a slightly lower transmit power at SINR above 30dB, the proposed learning-based model's performance is within $96\%-98\%$ of the optimization-based solution.

\subsection{Performance Evaluation of Robust SLP-DNet}\label{robust_performance}
In this subsection, we evaluate the performance of the robust SLP-DNet against the robust SLP optimization-based and conventional precoding algorithms.\par
\begin{figure}[!t]
  \includegraphics[width=2.85in,height=2.5in]{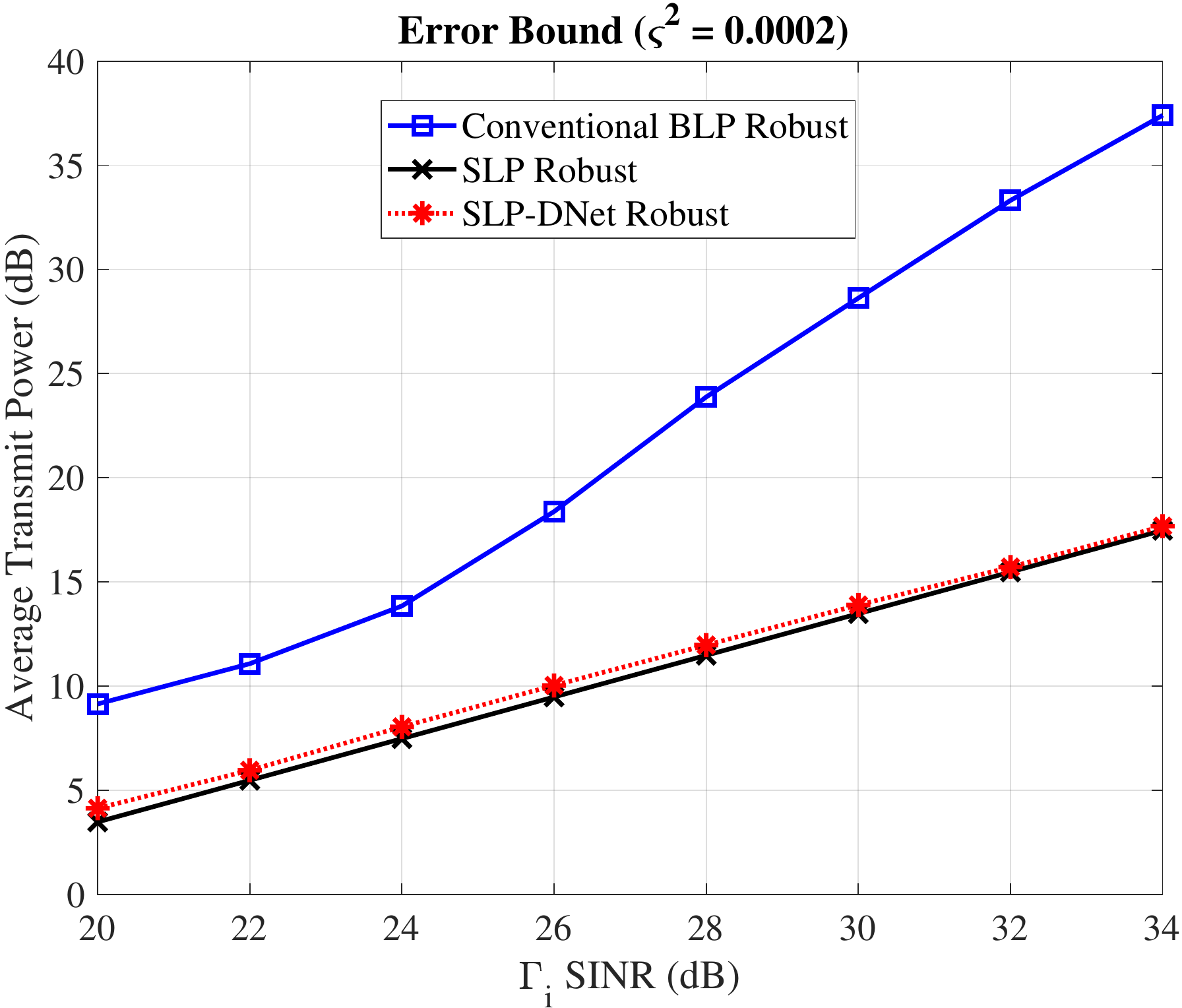}%
\caption{Transmit Power vs SINR averaged over 2000 test samples for robust conventional, SLP optimization-based and SLP-DNet solutions with $N_{t}=4$, $K=4$ and $\varsigma^{2}=0.0002$.}
\label{fig:POWER_vs_SINR_SQ_Robust}
\end{figure}
\begin{figure}[!t]
  \includegraphics[width=2.85in,height=2.5in]{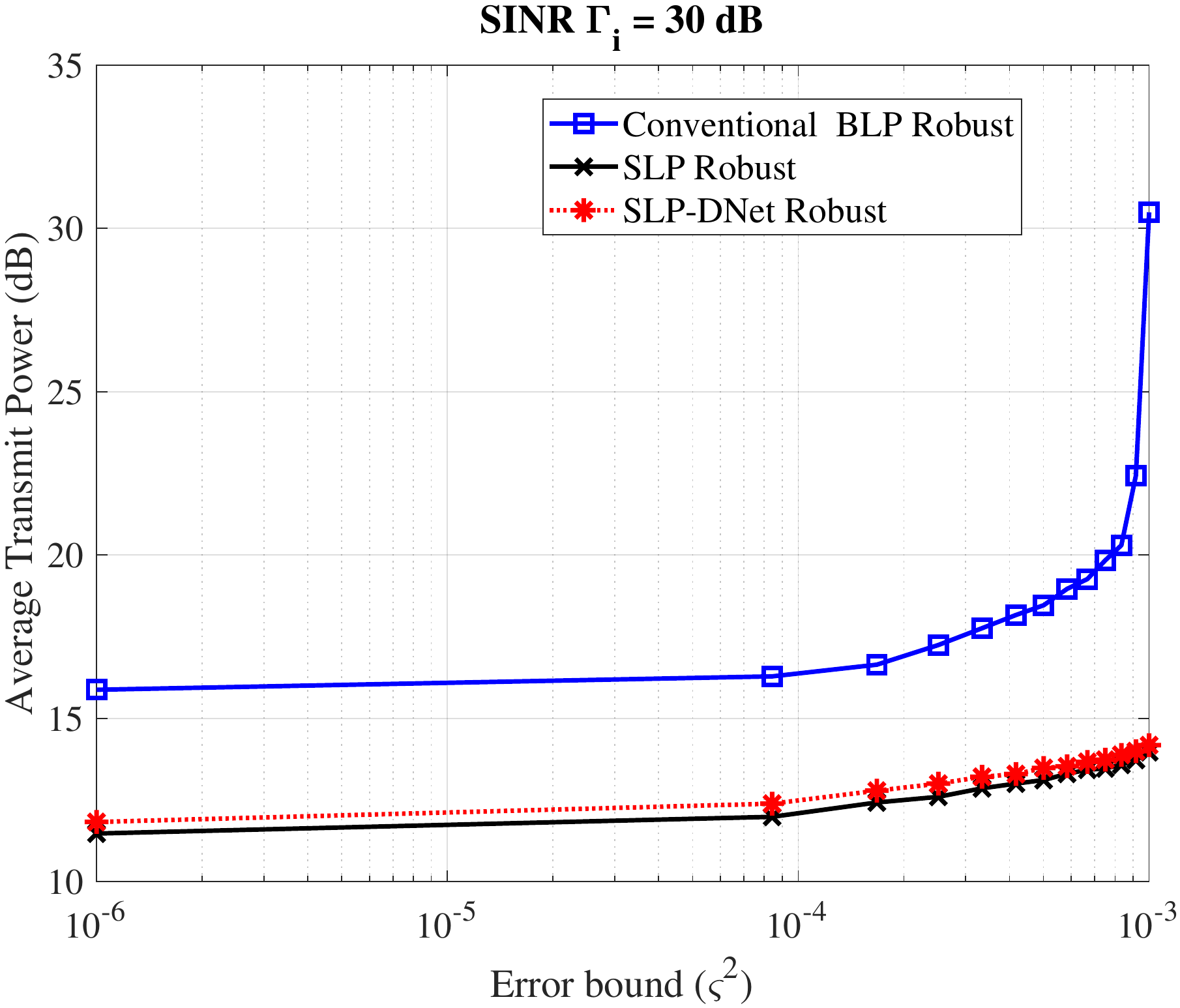}%
\caption{Transmit Power vs Error-bound for robust conventional BLP, SLP optimization-based and SLP-DNet solutions with $N_{t}=4$, $K=4$.}
\label{fig:POWER_vs_ERRORBOUND}
\end{figure}

Figs. \ref{fig:POWER_vs_SINR_SQ_Robust} and \ref{fig:POWER_vs_ERRORBOUND} compare the performance of the proposed robust SLP-DNet with the traditional robust block-level precoder \cite{zheng2008robust} and robust SLP precoder \cite{masouros2015exploiting} for the $4\times4$ MISO system evaluated at $\varsigma^{2}={10}^{-4}$. For simplicity, we use QPSK modulation scheme. Fig. \ref{fig:POWER_vs_SINR_SQ_Robust} depicts how the average transmit power increases with the SINR thresholds, for CSI error bounds $\varsigma^{2}={10}^{-4}$. The SLP optimization-based precoding scheme is observed to show a significant power savings of more than 60\% compared to the conventional optimization solution. Similarly, the proposed unsupervised learning-based precoder portrays a similar transmit power reduction trend. They show considerable power savings of $40\%-58\%$ against the conventional BLP.\par 
Furthermore, we investigate the effect of the CSI error bounds on the transmit power at 30dB. Fig. \ref{fig:POWER_vs_ERRORBOUND} depicts the transmit power variation with increasing CSI error bounds. Moreover, a significant increase in transmit power can be observed where the channel uncertainty lies within the region of CSI error bounds of $\varsigma^{2}=10^{-3}$. Interestingly, like the SLP optimization-based algorithm, the proposed SLP-DNet also shows a descent or moderate increase in transmit power by exploiting the constructive interference.


\begin{figure}[!t]
    \centering
    \includegraphics[width=3.25in,height=2.5in]{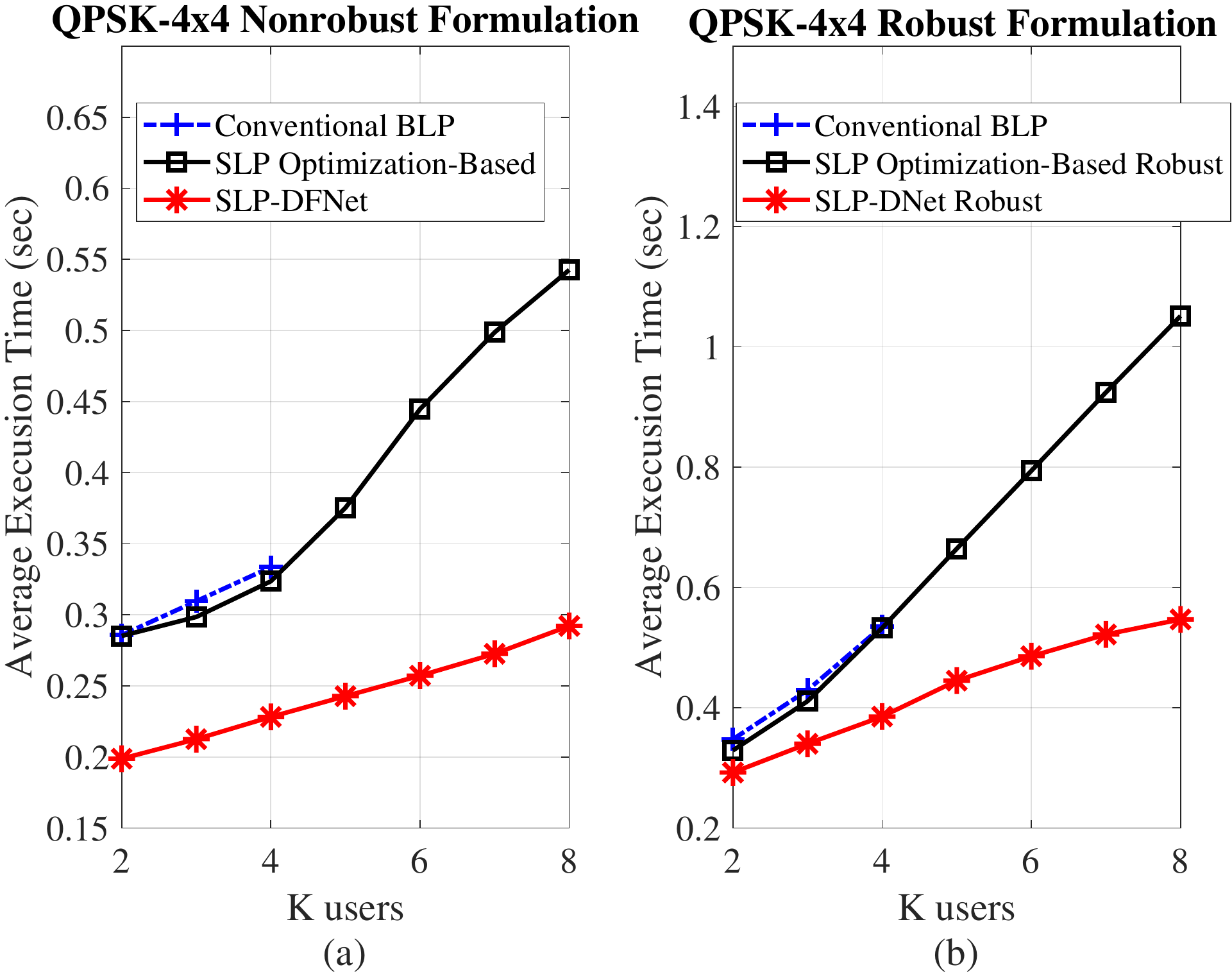}
    \caption{Comparison of average execution time per sample averaged over 200 test samples for conventional BLP, optimization-based and SLP-DNet solutions with $N_{t}=4$ and $K$ users $(2,\cdots,8)$.}
    \label{fig:Execusion_time}
\end{figure}

Figs. \ref{fig:Execusion_time}(a) and \ref{fig:Execusion_time}(b) depict the execution times for nonrobust and robust formulations. It can be seen that both SLP optimization-based algorithm and the proposed learning schemes are feasible for all sets of $N_{t}$ BS antenna and $K$ mobile users. However, for conventional BLP, the solution is only feasible for $N_{t} \geq K$.\par 
Fig. \ref{fig:Execusion_time}(a) shows the average execution time of the proposed unsupervised learning solutions per symbol averaged over 2000 test samples for nonrobust formulations. The SLP-DNet is observed to be significantly faster than the SLP optimization-based. For example, the theoretical complexity is polynomial order-3 and polynomial order-6.5 or order-7.5 for SLP-DNet and conventional methods, respectively. This is shown in the execution times, where there is a significantly steeper increase in run-time as the number of users increases. The decrease in computational cost is because the dominant operations involved in SLP-DNet at the inference are simple matrix-matrix or vector-matrix convolution. The same trend is also observed in the case of a robust channel scenario, as shown in Fig. \ref{fig:Execusion_time}(b). Therefore, the results in Figs. \ref{fig:Execusion_time}(a) and \ref{fig:Execusion_time}(b) demonstrate that the proposed unsupervised learning-based precoding solutions offer a good trade-off between the performance and computational complexity. Moreover, as per the results obtained, SLP-DNet's performance is within the range of $89\%-99\%$ of the optimal SLP optimization-based precoding solution. Thus, our proposals demonstrate a favorable tradeoff between the performance and the computational complexity involved. 

\section{Conclusion}\label{conclusion}
This paper proposes an unsupervised learning-based precoding framework for a multi-user downlink MISO system. The proposed learning technique exploits the constructive interference for the power minimization problem so that for given QoS constraints, the transmit power available for transmission is minimized. We use domain knowledge to design unsupervised learning architectures by unfolding the proximal interior point method barrier \textit{`log'} function. The proposed learning scheme is then extended to robust precoding designs with imperfect CSI bounded by CSI errors. We demonstrate that our proposal is computationally efficient and allows for feasible solutions to be obtained for problems where traditional numerical optimization like IPM and brute-force maximum likelihood solvers would not converge or would be prohibitively costly.

\begin{appendices}
\section{Proximity operator barrier for Robust SLP}\label{appendix} 
For every transmit precoding vector $\mathbf{w}_{2}\in\mathbb{R}^{2N_{t}\times1}$, the proximity operator of the barrier $\gamma\mu{B}_{\text{robust}}(\mathbf{w}_{2})$ is given by 
\begin{equation}\label{robusr_prox_opr_final}
    \Phi_{\text{rb}}(\mathbf{w}_{2},\gamma,\mu)=\frac{2\Gamma{v}_{0}\text{tan}^{2}{\phi}-X(\mathbf{w}_{2},\gamma,\mu)^{2}}{2\Gamma{v}_{0}\text{tan}^{2}{\phi}-X(\mathbf{w}_{2},\gamma,\mu)^{2}+2\gamma\mu}{\mathbf{w}_{2}}
\end{equation}
where $X(\mathbf{w}_{2},\gamma,\mu)$ is the unique solution of the cubic equation expressed as \cite{bertocchi2020deep}
\begin{multline}\label{robust_cubic_eqn}
   \resizebox{0.95\hsize}{!}{$\begin{array}{ll} {x}^{3}-\left(\left(\varsigma^{2}-\boldsymbol{\Lambda}^{T}\boldsymbol{\Lambda}\right)\norm{\mathbf{w}_{2}}_{2}+4\boldsymbol{\Lambda}^{T}\mathbf{w}_{2}\text{tan}{\phi}\sqrt{\Gamma{v}_{0}}\right){x}^{2}\\
    +\left(2\Gamma{v}_{0}\text{tan}^{2}{\phi}+2\gamma\mu\right){{x}}+\\
    2\Gamma{v}_{0}\text{tan}^{2}{\phi}\left(\left(\varsigma^{2}-\boldsymbol{\Lambda}^{T}\boldsymbol{\Lambda}\right)\norm{\mathbf{w}_{2}}_{2}+4\boldsymbol{\Lambda}^{T}\mathbf{w}_{2}\text{tan}{\phi}\sqrt{\Gamma{v}_{0}}\right)=0.\end{array}$}
\end{multline}
It can be observed that (\ref{robust_cubic_eqn}) is a cubic equation and can be solved analytically. In the final analysis, following similar steps as in (\ref{prox_opr_rl2})-(\ref{deriv_gamma_rl}), the robust deep-unfolded model is obtained by finding the Jacobean matrix of (\ref{robusr_prox_opr_final}) with respect to the optimization variable $\mathbf{w}_{2}$, and the derivatives with respect to the step-size $\gamma>0$ and the Lagrange multiplier associated with the inequality constraint $\mu>0$. We use similar concepts presented in subsection \ref{Prox_Operator_relaxed} to formulate the learning algorithm of the robust SLP as a series of sub-problems with respect to the combined effect of the two inequality constraints as follows
\begin{equation}\label{prox_robust}
    \begin{aligned}
    & \underset{\mathbf{w_{2} \in{\mathbb{R}}^{2N_{t}\times1}}}{\text{min}}
    & & {\norm{\mathbf{w}_{2}}}_{2}^{2}+\lambda{\mathbf{w}_{2}} +{\mu}{{B}_{\text{robust}}(\mathbf{w}_{2})}.
    \end{aligned}
\end{equation}
Similar to a nonrobust SLP-DNet, the update rule for every iteration is expressed as
\begin{equation}\label{beam_update_robust}
\mathbf{w}_{2}^{[r+1]}=\text{prox}_{\gamma^{[r]}{\mu}^{[r]}B_{\text{robust}}}\left(\mathbf{w}_{2}^{[r]}-\gamma^{[r]}\Delta{{D}_{\text{robust}}(\mathbf{w}_{2}^{[r]},\lambda^{[r]})}\right)
\end{equation}
where 
\begin{equation}\label{eq_constr_robust}
    {D}_{\text{robust}}(\mathbf{w}_{2}^{[r]},\lambda^{[r]})={\norm{\mathbf{w}_{2}}}_{2}^{2}+\lambda{\mathbf{w}_{2}}.
\end{equation}
 
\end{appendices}

\bibliographystyle{IEEEtran}
\bibliography{ref}

\end{document}